\documentclass{revtex4-2}

\usepackage{graphicx}
\usepackage{newtxtext}
\usepackage{newtxmath}
\usepackage{natbib}
\usepackage{booktabs}
\usepackage[inkscapelatex=false]{svg}
\usepackage[colorlinks=true,linkcolor=blue,allcolors=blue
]{hyperref}
\hypersetup{
    colorlinks = true,
    urlcolor   = red,
    citecolor  = blue,
}

\newcommand{\RomanNumeralCaps}[1]

\begin{document}

\title{Programmable ultrasonic fields enhance intracellular delivery in cell clusters}

\author{Subhas Nandy}
\affiliation{Micro Nano Bio Fluidics Unit, Department of Mechanical Engineering,
Indian Institute of Technology Madras, Chennai -- 600036, Tamil Nadu, India}

\author{Monica Manohar}
\affiliation{Micro Nano Bio Fluidics Unit, Department of Mechanical Engineering,
Indian Institute of Technology Madras, Chennai -- 600036, Tamil Nadu, India}

\author{Ashis K. Sen}
\email{ashis@iitm.ac.in}
\affiliation{Micro Nano Bio Fluidics Unit, Department of Mechanical Engineering,
Indian Institute of Technology Madras, Chennai -- 600036, Tamil Nadu, India}
\hrule

{\textsuperscript{*}To whom correspondence should be addressed. E-mail: ashis@iitm.ac.in}

\setlength{\arrayrulewidth}{0.5mm}
\renewcommand{\arraystretch}{1.5}
\setlength{\tabcolsep}{3pt}

{This manuscript was compiled on \today}

\fontsize{10}{11}\selectfont
\maketitle

\section*{Abstract}
Intracellular delivery of biomolecules remains a critical challenge in both basic cell biology and translational therapeutics. We introduce Programmable Acoustic Standing-wave Transfection (PAST), a microfluidic tool that leverages dynamically programmable ultrasonic fields to transiently permeabilize cell membranes and enhance biomolecular transport within cell clusters. By generating programmable acoustic potential landscapes, PAST drives cells through cycles of hydrodynamic and acoustic stresses that induce reversible pore formation, enabling diffusion-based delivery without chemical carriers or contrast agents. Experimental studies demonstrate controlled influx and efflux dynamics across multiple biomolecular species, with transport rates tunable via acoustic power, frequency modulation, and duty cycles. Theoretical scaling and numerical simulations reveal that membrane tension, pore energetics, and acoustic field distributions collectively govern transmembrane transport of biomolecules. Post-treatment assays confirm high cellular viability and sustained proliferation, underscoring the biocompatibility of the method. Remarkably, effective diffusivity estimates derived from model predictions closely match experimental transport timescales. Together, these findings establish PAST as a programmable, high-throughput, and non-invasive intracellular delivery platform, offering new opportunities for precision drug screening, gene editing, and mechanistic exploration of cellular membrane biophysics.\\
Keywords: Intracellular delivery, Ultrasonic waves, Dynamic acoustic fields.

\section{Introduction }
Efficient intracellular delivery is a cornerstone of biomedical engineering, underpinning applications in drug delivery, gene therapy, and personalized medicine \cite{zhai_drug_2024, stewart_vitro_2016, hur_microfluidic_2021}. Conventional chemical carriers risk immune activation and toxicity, while contact-based methods such as scrape loading and micropipette disruption are invasive and lack scalability \cite{schmiderer_efficient_2020, vandersarl_nanostraws_2012}. Physical force-mediated approaches, including hydrodynamic \cite{kang_intracellular_2020, kizer_hydroporator_2019}, mechanical \cite{uvizl_efficient_2021, jung_mechanoporation_2022}, electrical \cite{wu_micromotor-based_2021, ding_high-throughput_2017}, and optical \cite{prentice_membrane_2005, fan_efficient_2015} methods, offer alternatives, but each is limited by either equipment complexity, bio-incompatibility, or loss of cell viability. Ultrasound-based poration provides a non-invasive, biocompatible strategy, yet microbubble-assisted sonoporation suffers from inconsistent outcomes, membrane damage, and lack of control \cite{qin_mechanistic_2018, pereno_microstreaming_2020, tu_ultrasound-mediated_2022}, while contrast agent-free approaches often demand adherent cells or specialized GHz resonators \cite{kim_acoustofluidic_2022, ramesan_acoustically-mediated_2018, zhang_hypersonic_2017, guo_controllable_2021, shen_cell_2024}, restricting scalability for suspended populations. Bulk-wave Acoustofluidics offers an attractive framework to address these challenges by leveraging acoustic radiation forces (ARF) and acoustic streaming flows (ASF) for label-free manipulation of cells. Existing ultrasonic devices, however, often rely on multiple transducers and/or rigid frequency constraints \cite{courtney_dexterous_2013, courtney_independent_2014}, limiting programmability of acoustic potential landscapes and precise control of suspended cell aggregates. This gap in dynamic, bubble-free acoustic poration methods has restricted their broader utility in intracellular delivery applications. Here, we introduce a Programmable Acoustic Standing-wave Transfection (PAST) strategy, which dynamically reconfigures ultrasonic fields through frequency modulation in a single-transducer actuated microcavity. This approach enables spatiotemporal manipulation of cell clusters while exposing them to synergistic ARF- and ASF-mediated stresses, transiently permeabilizing membranes and enhancing biomolecular transport without any cavitation or microjetting-induced mechanisms. Theoretical and numerical analyses elucidate the physical mechanisms underlying poration, establishing PAST as a biocompatible, label-free, and scalable platform for intracellular delivery. 

\begin{figure*}[t]
    \centering
    \includegraphics[width=1\linewidth]{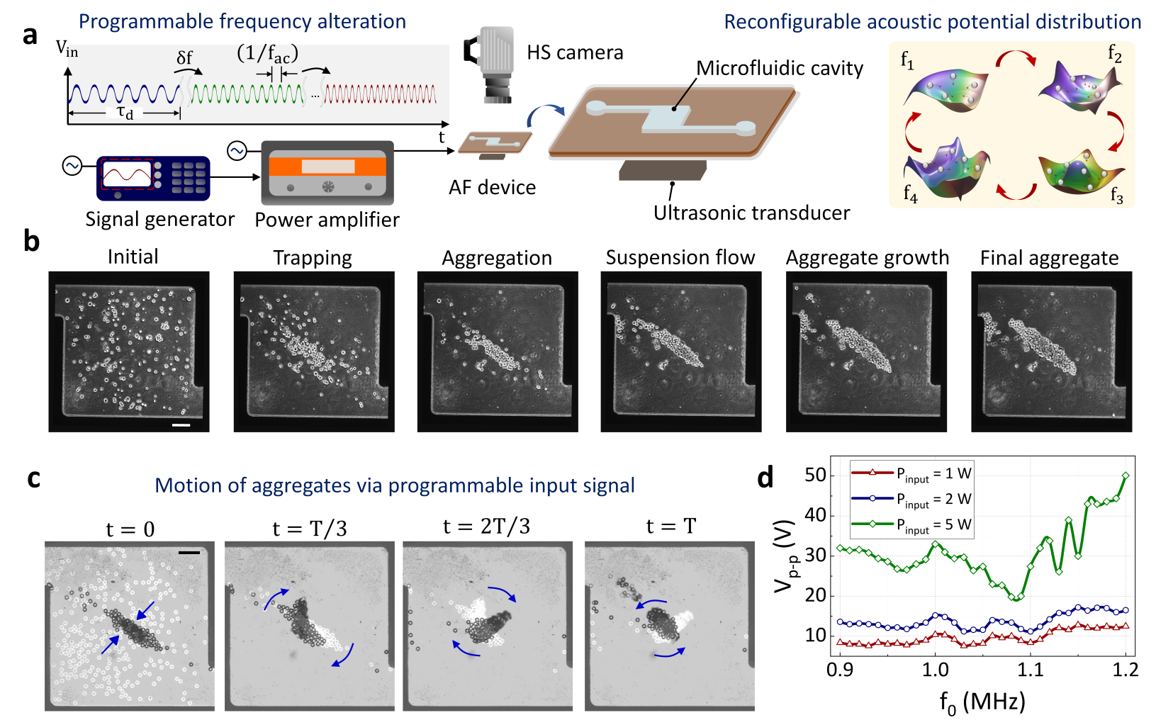}
    \caption{\textbf{Programmable Acoustic Standing-wave (PAS) and dynamics of bioparticle aggregates. }a. PAS setup with programmable frequency modulation generating reconfigurable acoustic potential landscapes. b. Experimental images showing cell trapping, aggregate nucleation, and growth into a desired cluster size. c. Aggregate motion under dynamic frequency modulation, illustrating translation, rotation, and morphological reconfiguration over a single frequency cycle (T). d. Measured peak-to-peak driving voltage across the acoustofluidic device as a function of actuation frequency for three acoustic powers.Scale bar represents 100 $\mu$m.}
    \label{fig1}
\end{figure*}

\section{ Results }
\subsection{Programmable Acoustic Standing-wave (PAS) and dynamics of bioparticle aggregates}

Ultrasonic wave propagation in microfluidic cavities gives rise to fluctuations in physical quantities such as pressure, density, velocity, and temperature, typically on microsecond timescales in the low-MHz regime. Although these oscillations average to zero within each acoustic cycle, their nonlinear interactions generate finite, time-averaged effects that persist over experimental timescales. Two dominant outcomes emerge: acoustic radiation forces (ARF), arising from scattering of acoustic waves at particle–fluid interfaces, and acoustic streaming-induced flows (ASF), which originate from viscous dissipation within boundary layers \cite{doinikov_acoustic_1997, sadhal_acoustofluidics_2012}. The balance between ARF and ASF provides a powerful means of manipulating suspended particles, allowing contact-free aggregation and manipulation within confined microscale environments \cite{levario-diaz_effect_2020}. The novelty of the programmable acoustic standing-wave (PAS) platform, lies in its ability to achieve dynamically reconfigurable acoustic potential landscapes. Rather than relying on fixed resonant conditions to generate static pressure nodes and antinodes, the PAS setup enables external modulation of the actuation frequency in a programmable manner. Because the acoustic propagation vector in a fluid-filled cavity is strongly frequency-dependent, periodic modulation of frequency results in evolving potential minima that drive suspended bioparticles along spatiotemporally varying trajectories (Fig. \ref{fig1}a) \cite{nandy_tunable_2025}. This reconfiguration is achieved by tuning two user-defined parameters: the frequency step size ($\Delta f$) and the dwell time ($\tau_d$) at each step. As aggregates track these evolving acoustic minima, they are exposed to combined hydrodynamic and ultrasound-induced stresses that transiently permeabilize cell membranes. Importantly, this mechanism avoids the complexities of ultrasound contrast agent (UCA)-based sonoporation \cite{fan_spatiotemporally_2012, qin_mechanistic_2018} and eliminates the need for multi-element arrays or complex electronics, relying instead on a single piezoelectric element and a programmable signal generator \cite{yang_harmonic_2022, dai_acoustic_2025, courtney_independent_2014}. To probe ultrasound-mediated permeabilization, adherent cells grown to $\sim 70 \%$ confluency are enzymatically detached, resuspended in medium supplemented with biomolecular cargo, and injected into the microcavity via syringe pumps. Resonant actuation at $f_r$ = 0.950 MHz generates ARF that traps cells within the cavity, while continued fluid inflow supplies additional cells to the aggregate until the desired cluster size is reached (Fig. \ref{fig1}b; Supplementary Fig. S3 and Supplementary Movie 1). The trapped aggregate is then subjected to programmed frequency sweeps, each defined by $\Delta f$ and $\tau_d$, with on/off periods of $\tau_{on}$ = 30 s and $\tau_{off}$ = 90 s to mitigate thermal buildup. The aggregate size can be further tuned by varying flow duration or initial seeding density. PAS operation is first validated using 20 $\mu$m polystyrene microparticles. Aggregates formed at resonance are driven by dynamically modulated input signals, as visualized in sequential micrographs captured during one sweep cycle (Fig.\ref{fig1}c and Supplementary Movie 2). Superimposed, color-inverted images highlight displacement between pre- and post-switch configurations, confirming controlled aggregate motion while maintaining structural integrity. Ultrasonic manipulation of biological cells is then experimentally validated by exposing acoustically aggregated HeLa cell clusters to dynamic reconfiguration of acoustic fields (Supplementary Movie 3). Electrical measurements reveal that at input power $P_{in}$ $\leq$ 2.0 W, the peak-to-peak voltage remains $\leq$ 20 V, a regime consistent with stable operation. At higher input powers ($P_{in}$ $\approx$ 5.0 W), $V_{p-p}$ increases to $\geq$ 40 V (Fig.\ref{fig1}d), conditions that risk excessive heating and impaired device performance. All subsequent experiments are therefore conducted at $P_{in}$ $\approx$ 2.0 W, ensuring biocompatibility as verified by post-ultrasound cell viability assays.

\begin{figure*}[h!]
    \centering
    \includegraphics[width=1\linewidth]{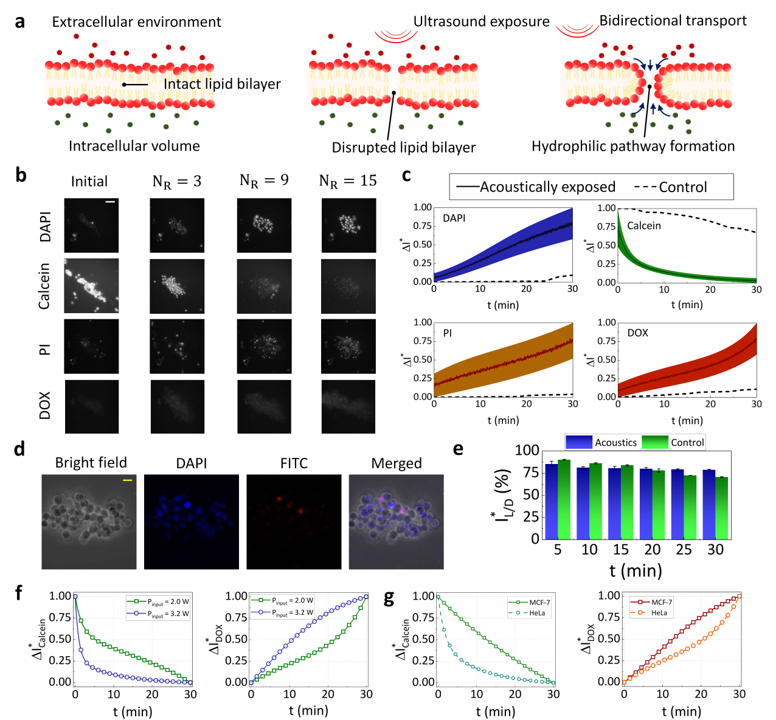}
    \caption{\textbf{Intracellular transport under Programmable Acoustic Standing-wave Transfection (PAST).}a. Schematic of membrane permeabilization under external stresses: disruption of the lipid bilayer initiates hydrophobic pores, followed by lipid head reorientation into hydrophilic pores permitting biomolecular transport in aqueous media. b. Fluorescence images showing intracellular uptake of different cargos after multiple frequency cycles ($N_R$); Scale bar, 100 $\mu$m. c. Temporal evolution of normalized fluorescence intensity ratio ($\Delta I^*$) over experimental timescales ($N_R$=15), indicating continuous uptake/efflux of biomolecules. d. Confocal images of acoustically treated HeLa aggregates stained with live–dead assay: blue = nuclei of all cells, red = cells with permanently compromised membranes; Scale bar, 20 $\mu$m. e. Time evolution of normalized fluorescence intensity ratio ($\Delta I_{L/D}^*$) of treated vs control cells over 30 min post-exposure; Scale bar, 20 $\mu$m. f. Evolution of $\Delta I^*$ for calcein and doxorubicin uptake in aggregates subjected to two input powers. g. Evolution of $\Delta I^*$  for calcein and doxorubicin in HeLa vs MCF-7 aggregates.}
    \label{fig2}
\end{figure*}

\subsection{Intracellular Transport under Programmable Acoustic Fields (PAST)}
The plasma membrane of biological cells acts as a selectively permeable barrier, regulating the passage of exogenous biomolecules \cite{cunill-semanat_spontaneous_2019}. During transmembrane transport, transient disruptions of the lipid bilayer create hydrophobic defects, wherein the lipid tails form the walls of nanoscale pores. These defects are inherently thermodynamically unstable and close spontaneously under equilibrium conditions. However, under external field-mediated stresses, the energetics of pore formation can be altered, enabling lipid molecules to reorient along the pore walls, forming hydrophilic linings that facilitate passage of biomolecules suspended in aqueous solvents \cite{cunill-semanat_spontaneous_2019, akimov_pore_2017}. This process is schematically illustrated in Fig. \ref{fig2}a. To experimentally validate intracellular delivery, we first studied three biomolecular species with distinct membrane permeation properties: a nucleic acid staining dye, DAPI; a live-cell tracking dye, Calcein-AM; and a membrane-impermeant dye, propidium iodide (PI). Fluorescence imaging was performed using filter sets appropriate to each fluorophore (Methods). For quantitative assessment, a non-dimensional fluorescence intensity ratio was defined as $\Delta I^{*} = (I(t) - I(0))/(I(t_m) - I(0))$, where $I(0)$ and $I(t_m)$ correspond to initial and final (after frequency sweep cycles, $N_R$=15) fluorescence intensities under dynamically varying ultrasonic exposure, and I(t) denotes intermediate values. Evolution of $\Delta I^*$ provides a direct measure of transmembrane transport under ultrasonic-mediated stresses. Experimental observations revealed rapid permeation and nuclear localization of DAPI, with fluorescence intensity progressively increasing over time (Fig.\ref{fig2}b, Supplementary Fig. S4 and Supplementary Movie 4). PI, typically membrane-impermeant, exhibited enhanced permeation, providing direct evidence of transient pore formation. In contrast, Calcein-AM, which passively permeates and is converted to fluorescent calcein intracellularly, exhibited efflux from treated cells, indicating bidirectional transport in response to membrane perforation. The combined uptake and efflux profiles, plotted in Fig. \ref{fig2}(c), highlight the global and nonspecific nature of membrane permeabilization induced by dynamic acoustic fields, confirming that large cell populations can be simultaneously targeted, which is advantageous for high-throughput studies (Supplementary Movie 5). Clinically relevant drug delivery was probed using doxorubicin, an anthracycline therapeutic that fluoresces upon intercalation with nuclear DNA/RNA \cite{aminipour_passive_2020}. Nondimensional fluorescence intensity measurements confirmed rapid nuclear uptake, with time-dependent profiles over repeated frequency sweep cycles (Fig.\ref{fig2}b, \ref{fig2}c). All experiments were repeated across independent trials (n = 10), with mean fluorescence evolution represented as solid lines and standard deviation bands capturing cell-to-cell heterogeneity, differences in aggregate confluency, and minor out-of-plane displacements. Supplementary Fig. S4 further demonstrates the non-specificity of PAST-mediated permeabilization. Cell viability was assessed using live/dead staining, with $\Delta I_{L/D}^*  = (1 - I_d/I_t) \times 100$, where $I_t$ and $I_d$ denote fluorescence intensities of all cells and dead cells, respectively. Evolution over 30 min post-exposure showed no significant difference between treated and control cells (Fig. \ref{fig2}d,e), establishing short-term biocompatibility under dynamically varying frequencies. The effect of acoustic driving power was systematically investigated at $P_{in}$ = 1.2 W, 2.0 W, and 3.2 W (Supplementary Fig. S5). Low-power excitation (1.2 W) failed to permeabilize membranes effectively, as evidenced by negligible calcein efflux and doxorubicin uptake. Increasing $P_{in}$ to 2.0 W enhanced transport rates, achieving saturation after $\sim$15 frequency sweep cycles, whereas 3.2 W accelerated transport, with saturation after $\sim$10 cycles (Fig. \ref{fig2}f), highlighting the tunability of PAST for controlling intracellular delivery. Finally, intracellular delivery was compared across HeLa and MCF-7 cell lines. Fluorescence evolution profiles revealed faster doxorubicin influx and slower calcein efflux in MCF-7 cells (Fig. \ref{fig2}g), likely due to differences in membrane potential and biophysical properties \cite{yang_membrane_2013, aminipour_passive_2020} (Supplementary Fig. S6 and Supplementary Note 1). These observations demonstrate the capability of dynamically altering acoustic fields to achieve enhanced, tunable intracellular transport of diverse biomolecules, including therapeutically relevant agents, while maintaining high cell viability.

\begin{figure*}[h!]
    \centering
    \includegraphics[width=1\linewidth]{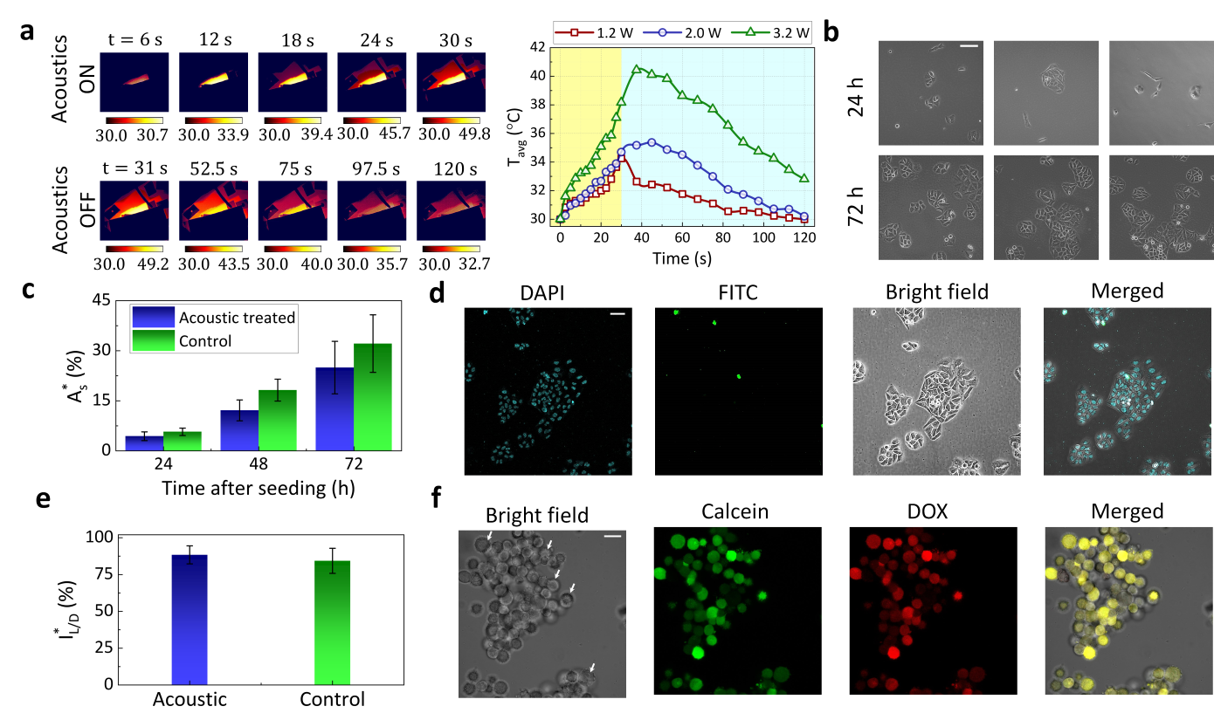}
    \caption{\textbf{Biocompatibility and Long-Term Cell Viability Post-PAST }a. Infrared imaging of temperature rise in the acoustofluidic device during frequency sweep cycles, figure shows average device temperature over one cycle at three input powers; background shading denotes actuation (30 s, 0.900–1.200 MHz) and resting (90 s) phases. b. Bright-field images of ultrasonically treated HeLa cells seeded on poly-L-lysine–coated dishes, monitored every 24 h; scale bar, 100 $\mu$m. c. Temporal evolution of normalized seeded cell area over 72 h compared with untreated controls. d. Confocal live/dead staining of proliferated cells after 72 h; scale bar, 100 $\mu$m. e. Normalized fluorescence intensity ratio for treated versus control cells, showing only marginal deviation and confirming sustained viability and biocompatibility. f. Confocal images of acoustically treated HeLa aggregates with calcein-AM and doxorubicin, demonstrating intracellular uptake and retention of doxorubicin with residual calcein signals after 72 h; scale bar, 20 $\mu$m. Inset arrows demarcate bleb formation on membrane surface.}
    \label{fig3}
\end{figure*}

\subsection{Biocompatibility and Long-Term Cell Viability Post-PAST}
While short-term cellular viability is often retained immediately following ultrasonic exposure, prior studies have reported that long-term viability and proliferation may be compromised due to perturbations of the cytoskeletal architecture and impaired restoration of cellular homeostasis \cite{duan_sonoporation_2021, wang_membrane_2012}. Therefore, evaluating biocompatibility is crucial to validate the applicability of this methodology for lab-on-chip drug screening and high-throughput intracellular delivery. To quantify the thermal impact of our programmable acoustic standing-wave (PAST) methodology, infrared imaging was employed to monitor the temperature evolution of the acoustofluidic device (Methods). Fig. \ref{fig3}a presents time-lapse snapshots during the actuation period ($\tau_{on}$ = 30 s) and resting period ($\tau_{off}$ = 90 s), along with the averaged temperature profiles for three acoustic driving powers: $P_{in}$ = 1.2 W, 2.0 W, and 3.2 W. At the highest driving power (3.2 W), device temperatures increased by $\sim 10^o$C and did not fully return to baseline during the resting period, potentially compromising cellular viability and throughput (Supplementary Fig. S7 and Supplementary Movie 6). Conversely, low-power excitation (1.2 W) maintained acceptable temperatures but failed to generate sufficient stresses for membrane permeabilization. Consequently, a driving power of $P_{in}$ = 2.0 W was selected for all subsequent experiments, balancing efficacy and biocompatibility (Supplementary Movie 7). Post-ultrasound viability was assessed by collecting acoustically treated cell aggregates via syringe-driven fluid flow into sterile, poly-L-lysine–coated Petri dishes containing fresh culture medium. The dishes were incubated under standard conditions (5$\% CO_2$, $37^o$C) and monitored periodically over 72 h. As shown in Fig. \ref{fig3}(b), treated cells exhibited normal adherence and healthy proliferation. Quantification using normalized seeded cell area, $A_s = (A_{cell}/A_{FOV}) \times 100$, revealed monotonically increasing trends over the observation period (Fig. \ref{fig3}c), with only marginal reductions relative to untreated controls, likely due to transient actin cytoskeleton disruption or membrane depolarization \cite{samandari_ultrasound_2017, qin_sonoporation-induced_2014}. Long-term viability was further confirmed using a live/dead fluorescence assay (Sigma Aldrich). Confocal microscopy images revealed blue DAPI-stained nuclei of all cells and green FITC-stained dead cells (Fig. \ref{fig3}d). Non-dimensional fluorescence intensity ratios, $\Delta I_{L/D}^*$, indicated negligible differences between treated and control populations after 72 h (Fig. \ref{fig3}e and Supplementary Fig. S8), confirming successful membrane resealing and restoration of cellular homeostasis. To validate intracellular retention of therapeutic agents, doxorubicin-loaded cells ($2 \times 10^6$ cells/mL) were exposed to 15 frequency sweep cycles at $P_{in}$ = 2.0 W, then washed and incubated in fresh medium for 24 h. Confocal imaging at 60$\times$ magnification revealed colocalization of doxorubicin fluorescence (TRITC filter, red) within the cytoplasm, along with residual calcein (FITC, green) (Fig. \ref{fig3}f and Supplementary Fig. S9). Notably, protrusion-like structures (indicated by arrows) on the cell membrane, consistent with blebs, were observed, suggesting transient ultrasonic-induced actin cytoskeleton remodeling \cite{gores_plasma_1990, leow_membrane_2015} (Supplementary Fig. S10). Cells exhibiting red fluorescence retained their spherical morphology, indicative of successful intracellular localization and doxorubicin-mediated cytotoxicity, while healthy cells maintained adherence and spreading on the lysine-coated surface. Overall, these results demonstrate that PAST-mediated intracellular delivery achieves high efficacy of cargo internalization while preserving short- and long-term cell viability, with careful control of acoustic driving parameters enabling a balance between permeabilization and biocompatibility.

\begin{figure*}[h!]
    \centering
    \includegraphics[width=1\linewidth]{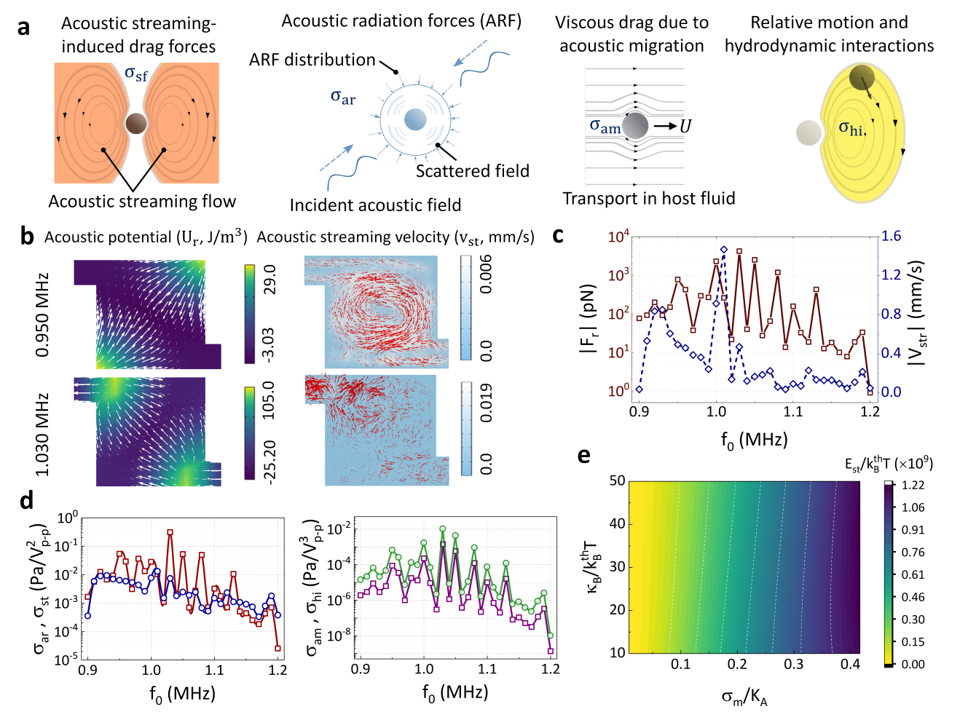}
    \caption{\textbf{Mechanisms of Ultrasound-Induced Cell Permeabilization} a. Schematic of stress-imposing mechanisms in the PAST setup, including acoustic radiation, streaming-induced drag, transport-related drag, and hydrodynamic interactions. b. Simulated acoustic potential and streaming fields at two actuation frequencies (0.950 and 1.030 MHz), with arrows indicating acoustic radiation force (ARF) (white) and acoustic streaming velocity (ASV) (red). Frequency modulation induces periodic changes in the force landscape. c. Predicted non-monotonic dependence of ARF and ASV magnitudes on frequency. d. Frequency-dependent stresses from four mechanisms, normalized to input voltage. e. Stored membrane strain energy under ultrasonic stresses far exceeds thermal fluctuation scales, facilitating pore formation.}
    \label{fig4}
\end{figure*}

\begin{figure*}[h!]
    \centering
    \includegraphics[width=1\linewidth]{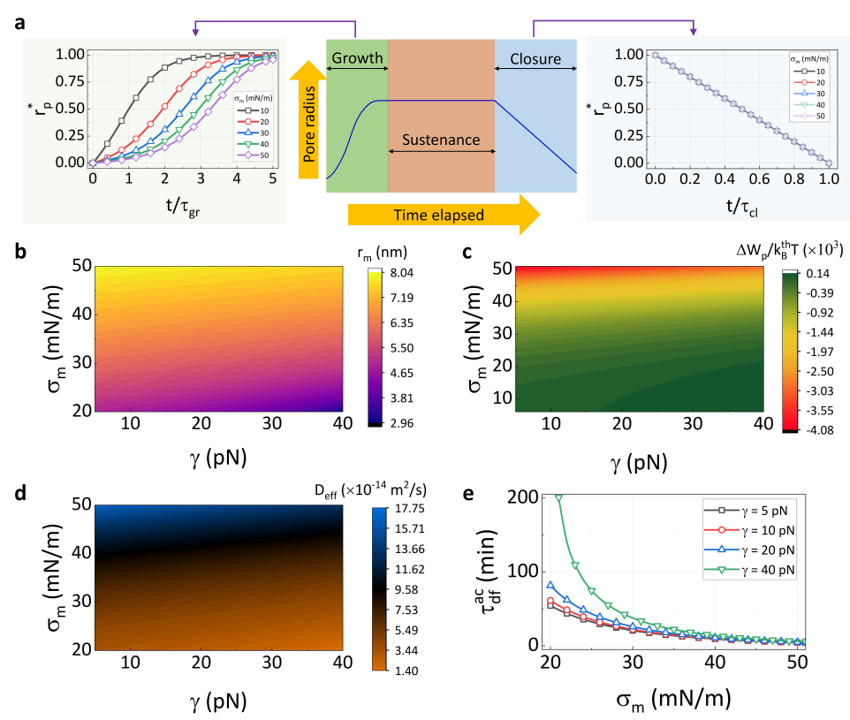}
    \caption{\textbf{Pore formation dynamics and enhanced diffusion across permeabilized cell membranes. } a. Schematic of the stages in pore lifetime, with plots showing pore growth and closure over non-dimensionalized time, b. Predicted variation of maximum pore radius with edge tension ($\gamma$) and applied membrane tension ($\sigma_m$), c. Energy barrier for pore formation, normalized to the thermal fluctuation scale ($k_B^{th} T$), as a function of edge tension and membrane tension, d. Theoretical predictions of enhanced diffusion coefficients for doxorubicin transport across membranes as a function of edge tension and membrane tension, e. Predicted timescales for doxorubicin transmembrane transport versus applied membrane tension at different edge tension values.}
    \label{fig5}
\end{figure*}

\subsection{Mechanisms of Ultrasound-Induced Cell Permeabilization}
Permeabilization of cell membranes under dynamically modulated acoustic fields arises from interdependent mechanisms driven by ultrasound-induced radiation and hydrodynamic stresses. Numerical simulations and theoretical scaling elucidate these effects. Material property contrasts between cells and fluid lead to acoustic scattering, transferring momentum and generating primary acoustic radiation forces (ARF) \cite{doinikov_acoustic_1997}. Viscous dissipation at fluid-solid interfaces induces acoustic streaming flow (ASF) \cite{sadhal_acoustofluidics_2012}, producing drag forces on cells (Supplementary Movie 8). Temporal modulation of actuation frequency shifts acoustic potential minima, driving acoustophoretic migration, while relative motion and hydrodynamic interactions among cells within aggregates further contribute to membrane stresses. These mechanisms (and corresponding stresses: $\sigma_{ar}$,$\sigma_{sf}$, $\sigma_{am}$ and $\sigma_{hi}$) are illustrated in Fig. \ref{fig4}a. To quantitatively characterize these processes, numerical simulations with full electromechanical coupling of the piezoelectric actuator, solid substrates, and fluid medium are performed. Vibrational displacements computed from these simulations serve as sources for acoustic pressure and streaming velocity fields under varying actuation frequencies (Methods; Supplementary Information). Experimental validation of pressure amplitudes and patterns is achieved by tracking microparticle trajectories and aggregation, as reported previously \cite{nandy_tunable_2025}. Acoustic radiation forces and streaming velocity distributions at four representative actuation frequencies are presented in Fig. \ref{fig4}b and Supplementary Fig. S11, with the evolution of these parameters over the full operational frequency range shown in Fig. \ref{fig4}c. Theoretical scaling analyses of individual stress contributions as a function of frequency are performed using fundamental principles (Methods), and presented in Fig. \ref{fig4}d. As stress magnitudes are strongly dependent on acoustic driving power, values are normalized with the peak-to-peak driving voltage, $V_{p-p}$. Volume incompressibility of cells translates these external forces into an equivalent membrane tension according to the Young-Laplace relation \cite{isambert_understanding_1998} as: $\sigma_m = \Sigma_{eq} \times (R_{cell}/2)$, where $R_{cell}$ is the cell radius and $\Sigma_{eq}$ denotes the net equivalent stress.  Elastic strain energy stored within the cell membrane (Methods) is found to exceed the thermal fluctuation energy scale by several orders of magnitude, thereby facilitating the crossing of energy barriers and promoting pore formation (Fig. \ref{fig4}e). Together, acoustic radiation, streaming-induced drag, frequency-modulated acoustophoretic motion, and hydrodynamic interactions drive transient pore formation in cell membranes, thereby enabling controlled intracellular delivery under programmable acoustic fields.

\subsection{Dynamics of Pore Formation in the Cell Membrane}
The plasma membrane selectively regulates biomolecule transport while maintaining homeostasis. Pore formation is energetically controlled, with thermal fluctuations enabling stochastic nucleation under passive conditions \cite{akimov_pore_2017-1}. Dynamically modulated external fields, such as acoustic stresses, lower the energy barrier, enhancing pore formation probability and kinetics \cite{akimov_pore_2017}. A nucleated pore undergoes growth, sustenance, and closure (Fig. \ref{fig5}a). Pore dynamics result from the competition between membrane tension ($\sigma_m$) promoting expansion and edge tension ($\gamma$) driving closure; pores spontaneously close when $\sigma_m < \gamma/r$. Initial growth is linear with timescale $\tau_{nu} = \eta h_m/\gamma$, transitioning to a transcendental phase with $\tau_{gr} = \eta h_m/\sigma_m$. Membrane tension relaxes during expansion as $\sigma(r) = \sigma_m(1 - r^2/r_c^2)$ (Methods). The critical pore radius can be estimated by evaluating the balance between membrane stresses arising from thermal fluctuations and externally applied forces \cite{sandre_dynamics_1999}(see Methods).

Membrane disruption triggers repair mechanisms, including $Ca^{2+}$ influx and cytoskeletal remodeling, restoring homeostasis and preserving cell viability \cite{zhou_effects_2008, jia_spatiotemporal_2022}. These processes restore homeostasis, prevent intracellular leakage, and maintain cell viability, consistent with our experimental observations. Upon removal of external stress, pores close over a timescale $\tau_{cl} = (2\eta h_m/\gamma)(r_c - r_i)$, where $\eta$ is the effective membrane viscosity, $h_m$ the membrane thickness, $\gamma$ is the edge tension, $r_c$ the critical radius and $r_i$ the initial nucleation radius. For typical membrane parameters, $\tau_{cl} \sim $10 s, much shorter than the interval between frequency sweeps ($\tau_{off} \sim$90 s). This allows full pore closure between cycles, preventing residual stresses and justifying cycle-wise independence in theoretical analyses. A non-dimensional pore radius, $r_p^*=(r_c-r)/(r_c-r_i)$, is defined to track pore evolution, with time scales adapted to the pore lifetime stage (Fig. \ref{fig5}a). Pore growth exhibits nonlinear dependence on membrane properties, including edge tension, bending modulus ($\kappa_B$), and area expansion modulus ($K_A$). Higher membrane tension reduces nucleation radius and increases critical radius, extending growth to $\sim 5 \tau_{gr}$. In contrast, closure dynamics collapse onto a single master curve, dominated by edge tension. Maximum pore radius (Fig. \ref{fig5}b) scales nonlinearly with membrane tension and edge tension; higher edge tension impedes growth of pores and limits maximum pore radius while increasing membrane tension energetically favors the formation of larger pores. Low acoustic power ($P_{in}$ = 1.2 W) produces small pores and minimal transport, while higher powers enhance pore size and transport (Extended data Fig. \ref{fig_ext3}). Excessive power risks irreversible poration and heating, compromising cell viability. Work done in pore formation at maximum radius is calculated (Methods) versus membrane and edge tension (Fig. \ref{fig5}c). Pore formation probability follows Boltzmann statistics, $\sim exp(-\Delta W_p/k_B^{th } T)$, where $\Delta W_p$ is the work needed to form a pore of a given radius, normalized by the thermal energy scale \cite{ma_simulation_2025, akimov_pore_2017}. Low external tension maintains a positive energy barrier, favoring closure, whereas increased membrane tension inverts the barrier, promoting spontaneous pore formation (Methods). Conversely, higher edge tension increases the barrier, driving pore closure. Estimating enhanced transport rates through perforated membranes requires knowledge of the effective pore area. While pore formation is inherently stochastic \cite{akimov_pore_2017-1}, an order-of-magnitude estimate of the number of pores can be obtained by balancing effective membrane tension and edge tension, $\sigma(r)-(\gamma/r) = 0$, corresponding to $r = r_m$ (Methods). Such theoretical scaling is essential for subsequent computation of enhanced diffusivity across the permeabilized cell membrane.

\subsection{Enhanced Diffusion across Permeabilized Cell Membranes}
Cell membrane permeabilization under dynamically modulated acoustic fields enhances biomolecular transport across the plasma membrane. Hindered diffusion theory models this effect, treating pores as cylindrical conduits spanning the membrane (Methods and Supplementary Information). Transport is governed by molecular diffusion and hydrodynamic interactions with pore walls, with retardation captured via an effective drag coefficient dependent on the solute-to-pore size ratio \cite{deen_hindered_1987}. The effective diffusivities for individual biomolecular species passing through a single pore are theoretically estimated using this approach. Normalization of these values relative to free-space diffusivity reveals stronger retardation effects for larger molecules, consistent with the enhanced hydrodynamic resistance encountered within confined pores (Supplementary Fig. S12). Externally applied stresses, which increase membrane tension, consequently enlarge pore radii. Larger pores reduce hydrodynamic hindrance, enhancing diffusivity coefficients and promoting more efficient transport across individual pores. To account for the collective effect of multiple pores oriented arbitrarily across the membrane surface, the effective diffusivity through a porous membrane is computed by incorporating the area fraction occupied by pores on the membrane surface \cite{skvortsov_permeability_2023} (Methods). Theoretical estimates of effective diffusivity for transport of doxorubicin and other biomolecules as a function of applied membrane and edge tensions are presented in Fig. \ref{fig5}d and Supplementary Information. Experimental studies on passive diffusion of doxorubicin through lipid bilayers and biological cells report permeability values, which, when scaled by membrane thickness, yield diffusion coefficients on the order of $\sim 10^{-15} m^2/s$ \cite{aminipour_passive_2020, speelmans_transport_1994}. In the presence of permeabilizing acoustic fields, increased effective diffusivity is expected to reduce the characteristic time scales for transmembrane transport. Assuming uniform filling of the cellular volume post-permeation, and considering the spatial resolution of our imaging experiments, the effective time scale for transport can be estimated (Methods). Time scales for transport of doxorubicin and other biomolecules, computed as a function of membrane tension and edge tension, are plotted in Fig. \ref{fig5}e and Supplementary Information. Molecular dynamics simulations indicate that a membrane tension of $\sigma_m \approx 38$ mN/m is sufficient to stabilize hydrophilic pores \cite{leontiadou_molecular_2004, krasovitski_intramembrane_2011}. At this reference tension, the theoretical transport time scale is $\sim$ 30 min, which aligns closely with experimental observations of doxorubicin fluorescence intensity evolution under dynamically modulated ultrasonic exposure (Fig. \ref{fig5}e). 

\section{Discussion}
Programmable Acoustic Standing-wave Transfection (PAST) establishes a new paradigm for non-invasive, high-throughput intracellular delivery by leveraging dynamically tunable ultrasonic fields to transiently permeabilize cell membranes within suspended clusters. Unlike conventional chemical or mechanical approaches that suffer from toxicity, invasiveness, or low scalability \cite{schmiderer_efficient_2020,vandersarl_nanostraws_2012,kang_intracellular_2020,kizer_hydroporator_2019,uvizl_efficient_2021,jung_mechanoporation_2022}, and unlike cavitation-based sonoporation, which introduces uncontrolled stresses and membrane damage \cite{ramesan_acoustically-mediated_2018,zhang_hypersonic_2017,guo_controllable_2021,tu_ultrasound-mediated_2022,shen_cell_2024}, PAST achieves precise, bubble-free modulation of acoustic potential landscapes. By integrating programmable frequency modulation within a single-transducer microcavity, this method enables spatiotemporal control over acoustic radiation forces (ARF) and acoustic streaming flows (ASF), which synergistically govern both aggregate motion and membrane poration.

The dynamic interplay between ARF and ASF generates temporally varying stress fields that collectively act on cell aggregates, inducing transient pores through localized membrane tension and viscous drag \cite{dai_acoustic_2025,cunill-semanat_spontaneous_2019,akimov_pore_2017,yang_membrane_2013}. Numerical simulations and theoretical scaling confirm that frequency-dependent shifts in acoustic pressure nodes and streaming vortices impose alternating compressive and shear stresses on the cell surface. These stresses raise the local membrane tension,$\sigma_m$, beyond the threshold for pore nucleation while remaining below cytotoxic limits. The stored elastic strain energy far exceeds the thermal fluctuation scale ($k_B T$), thereby reducing the energy barrier for pore formation \cite{ma_simulation_2025,leontiadou_molecular_2004}. As the actuation frequency is modulated, the spatially evolving acoustic field redistributes these stresses, producing repetitive cycles of pore opening and closure without inducing irreversible damage. This dynamic exposure underpins the observed reversible transport of both membrane-permeant and impermeant biomolecules.

Experimental fluorescence assays revealed efficient intracellular uptake of DAPI and doxorubicin, as well as bidirectional transport of Calcein-AM, confirming transient and nonspecific membrane permeabilization. The dependence of transport kinetics on input power demonstrates tunability of pore dynamics through acoustic parameters. Theoretical models predict that higher $\sigma_m$ and lower edge tension $\gamma$ increase the critical pore radius and lower the nucleation barrier, consistent with observed enhancement of transport rates under moderate acoustic power. Computed pore closure timescales ($\tau_{cl}\approx10 s$) remain shorter than the inter-sweep resting interval ($\tau_{off}\approx90 s$), ensuring full resealing between actuation cycles.

Enhanced diffusion across permeabilized membranes was modeled via hindered diffusion theory, accounting for solute–pore hydrodynamic interactions \cite{deen_hindered_1987,g_davidson_hydrodynamic_1988}. Estimated effective diffusivities for doxorubicin ($\sim 10^{-15}  m^2/s$) closely match experimental transport timescales ($\sim$ 30 min), validating the theoretical framework and confirming that acoustic modulation accelerates molecular transport without compromising integrity. Long-term culture studies further verified biocompatibility, as treated cells maintained normal adherence and proliferation over 72 h, consistent with transient cytoskeletal remodeling and homeostatic recovery \cite{zhou_effects_2008,jia_spatiotemporal_2022,sandre_dynamics_1999}.
Collectively, these results position PAST as a robust, scalable, and label-free platform for programmable intracellular delivery. By coupling tunable ultrasound fields with confined microfluidic geometries, PAST bridges the gap between static acoustofluidic trapping \cite{courtney_independent_2014,courtney_dexterous_2013,dai_acoustic_2025,nandy_tunable_2025} and invasive poration methods, offering unprecedented control over stress landscapes and membrane transport dynamics. Beyond intracellular delivery, the approach provides a versatile tool for probing mechanobiological responses to dynamic stresses, screening mechanosensitive therapeutics, and studying membrane repair pathways. The ability to dynamically encode acoustic fields introduces a new degree of freedom in acoustofluidics, transforming ultrasound from a passive manipulation tool into a programmable actuator for controlling intracellular transport.

\section{Materials and Methods}

\subsection{Acoustofluidic device fabrication and setup}
The device (Fig. \ref{fig1}a and Supplementary Fig. S1) consists of a 200 $\mu$m thick silicon substrate with a through-etched 750 $\mu$m square microcavity and 175 $\mu$m wide inlet-outlet channels, sandwiched between two 500 $\mu$m glass substrates. Structures were patterned in silicon using photolithography and deep reactive ion etching (DRIE), followed by anodic bonding to the glass layers. Fluidic access was enabled via 1 mm holes drilled in the glass. A 1 MHz piezoelectric transducer (Sparkler Ceramics, India) was bonded with epoxy to the bottom glass, offset from the microcavity to permit optical access. Experiments employed polystyrene microparticles (Sigma-Aldrich, India; 20 $\mu$m). Suspensions were prepared by diluting 100 $\mu$L of a $10\%$ v/v stock into 3 mL DI water, adding 2 $\mu$L Tween-20 to prevent premature microparticle agglomeration, followed by sonication for 15 min. The suspension was introduced into the cavity via a syringe pump (neMESYS, Cetoni, Germany). The transducer was driven with a sinusoidal signal (0.900–1.200 MHz) from a function generator (Rohde $\&$ Schwarz, Germany) amplified by a power amplifier (Rohde $\&$ Schwarz, Germany). Microparticle migration and patterning were visualized using a fluorescence microscope (Olympus, Japan) and recorded with a high-speed camera (Photron, UK; 250–1000 fps) via FastCAM Viewer 4. Voltage drop across the device was measured using an oscilloscope.

\subsection{Temperature measurement of acoustofluidic device} 
Dynamic temperature mapping was performed with an infrared camera (A6701, FLIR). Calibration was done by heating a reference plate (30–90 °C, 5 °C increments). The IR camera was positioned vertically with the device at the focal plane. Temperature evolution during frequency sweeps was captured at 30 fps. Pixel grayscale values were extracted using ImageJ and converted to temperature using calibration data and MATLAB routines. Average device temperature was obtained by surface-averaging over the device area.

\subsection{ Cell preparation and culture}
HeLa cells (NCCS, Pune, India) were stored at -80 °C, revived at 37 °C, and cultured in DMEM (Gibco) with 10$\%$ FBS (Himedia) and 1$\%$ Antibiotic-Antimycotic (Gibco) in a T-25 flask under 5$\%$ $CO_2$ at 37°C. Cells were passaged every three days. At $\sim 70\%$ confluence, they were washed with PBS (Himedia, pH 7.4), detached using 0.25$\%$ trypsin-EDTA (Gibco), centrifuged (1500 rpm, 5 min), resuspended, and counted. For fluorescence assays, cells were stained with 5 $\mu$M Calcein-AM (20 min), centrifuged, and resuspended in FluoroBrite DMEM (Gibco) at $2 \times 10^6$ cells/mL. Viability was assessed using ReadyProbesTM Cell Viability Imaging Kit (Invitrogen). After staining and treatment, cells were washed and imaged in FluoroBrite DMEM using fluorescence microscopy with DAPI (excitation/emission: 360/460 nm) and FITC/GFP (excitation/emission: 488/530 nm) filters. The influx trends of propidium iodide and doxorubicin into cells are visualized on TRITC filter (excitation/emission: 560/620 nm). For short-term viability experiments, observations were made periodically after every 5 mins. For evaluation of long-term viability, seeded cells were imaged periodically after every 12 h.

\subsection{Theoretical framework and numerical model}
\subsubsection{Acoustic field distributions}

Perturbation theory is used to resolve acoustic field distributions and time-averaged effects. Physical variables are decomposed into first-order, time-harmonic components (acoustic pressure and velocity in the fluid, and solid displacement) and second-order components, which contribute to time-averaged effects. The governing continuity and Navier–Stokes equations for the fluid, together with Cauchy’s equation for the solid, are solved order by order in the perturbation expansion \cite{nandy_tunable_2025}:

\begin{gather*}
      p = p_0 + \epsilon p_1 + \epsilon^2 p_2 + ... \\ 
    \rho = \rho_0 + \epsilon \rho_1 + \epsilon^2 \rho_2 + ... \\ 
    \vec{V} = \vec{V}_0 + \epsilon \vec{V}_1 + \epsilon^2 \vec{V}_2 + ...
\end{gather*}

Subscripts 0 denote the unperturbed fluid state, while subscripts 1 denote first-order acoustic variables oscillating at angular frequency $\omega$; higher orders follow similarly. The small parameter $\epsilon = |\vec{V}_1|/c_0$ represents the ratio of first-order particle velocity to sound speed. Substituting these variables into the governing fluid equations yields the first-order equations describing the fluid response to acoustic wave propagation,

\begin{equation}
    \frac{\partial\rho_1}{\partial t} + \nabla .(\rho_0 \vec{V}_1) = 0
\end{equation}
\begin{equation}
    \rho_0 \biggl( \frac{\partial \vec{V}_1}{\partial t} \biggr) = -\nabla p_1 + \mu \nabla ^2 \vec{V}_1 + \biggl( \mu_B + \frac{\mu}{3}\biggr) \nabla (\nabla . \vec{V}_1)
\end{equation}

Here, $\mu$ and $\mu_B$ denote the dynamic and bulk viscosities of the fluid, respectively. Assuming harmonic variation of the first-order variables, $q(\vec{r} ,t) = \tilde{q}(\vec{r})e^{-i \omega t}$, with $p_1 = c_0^2 \rho_1$ under adiabatic conditions at MHz frequencies, the governing equations reduce to the Helmholtz form for the first-order fields:

\begin{equation}
    \nabla^2 p_1 + \frac{\omega^2}{c_0^2} \bigl( 1 - i\Gamma_{ac} \bigr)p_1 = 0
\end{equation}
\begin{equation}
    \vec{V}_1 = \frac{-i}{\rho_0 \omega} \bigl( 1 - i\Gamma_{ac} \bigr) \nabla p_1
\end{equation}

Due to harmonic time dependence, the temporal term $e^{-i \omega t}$ is eliminated, and the first-order system response incorporates an effective acoustofluidic loss factor $\Gamma_{ac}$. This factor accounts for damping from viscous attenuation and boundary-layer formation at oscillating fluid–solid interfaces:

\begin{equation}
    \Gamma_{ac} = \frac{\omega}{2\rho_0 c_0^2} \biggl( \mu_B + \frac{\mu}{3} \biggr) + \frac{S}{4V}\sqrt{\frac{2\mu}{\rho_0 \omega}}
\end{equation}

Here, $S$ refers to the surface area of the fluid-solid interface on which the development of acoustic boundary layer is expected due to ultrasonic actuation, while $V$ refers to the volume of the fluid within the microfluidic system.

Elastic wave propagation in the solid substrates was evaluated by solving Cauchy’s equations:

\begin{equation}
    -\rho_{sl} \omega^2 (1 + i\Gamma_{sl})\vec{u}_{sl} = \nabla. \bar{\bar{\sigma}}_{sl}
\end{equation}

where, $\rho_{sl}$ is the density of the solid medium, $\vec{u}_{sl}$ refers to the displacement vector of the solid medium, $\bar{\bar{\sigma}}_{sl}$ refers to the stress tensor of the solid and $\Gamma_{sl}$ is the damping loss factor in the solid. The relationship between the stress in the elastic solid and the displacement of the solid is given by the constitutive relationship, $\bar{\bar{\sigma}}_{sl} =\bigl[ \bar{\bar{C}}_{sl}\bigr]. \bar{\bar{\varepsilon}}_{sl}$, where $\bigl[ \bar{\bar{C}}_{sl}\bigr]$ denotes the stiffness matrix of the elastic solid (characteristic of the nature of the material) and $\bar{\bar{\varepsilon}}_{sl}$ is the strain tensor of the solid which is related to the gradients of the displacement field of the solid.

Fluid–solid coupling was enforced by continuity of first-order velocity and stress across the interface:

\begin{equation}
    \vec{V}_{1,fl}.\vec{n} = -i \omega \vec{u}_{sl}.\vec{n}
\end{equation}
\begin{equation}
    -p_{fl} \vec{n} = \bar{\bar{\sigma}}_{sl}.\vec{n}
\end{equation}

where, $\vec{n}$ denotes the outward normal vector to the surface at the fluid-solid interface while the subscripts $fl$ and $sl$ refer to solid and fluid medium respectively. Free boundary conditions were applied at the solid–air interface , $\bar{\bar{\sigma}}_{sl}.\vec{n} = 0$.

Products of first-order fields generate second-order quantities, which yield finite non-zero values upon time-averaging: 

\begin{equation}
    \biggl\langle \frac{\partial \vec{\rho}_2}{\partial t}\biggr \rangle\ + \rho_0 .\langle \vec{V}_2 \rangle = -\nabla. \langle \rho_1 \vec{V}_1\rangle
\end{equation}

\begin{equation}
    \rho_0 \biggl\langle \frac{\partial \vec{V}_2}{\partial t}\biggr \rangle\ + \biggl\langle \rho_1\frac{\partial \vec{V}_1}{\partial t}\biggr \rangle +\rho_0 .\langle (\vec{V}_1.\nabla)\vec{V}_1 \rangle = -\nabla \langle p_2\rangle + \mu_0 \nabla^2 \langle \vec{V}_2 \rangle + \biggl( \mu_B + \frac{\mu_0}{3} \biggr) \nabla(\nabla . \langle \vec{V}_2\rangle)
\end{equation}
    
At vibrating fluid–solid interfaces, Stokes’ drift velocity was imposed as a boundary condition to enforce no-slip with the oscillating walls:

\begin{equation}
    \langle \vec{V}_{2} \rangle_{wall} = - \Bigl \langle \biggl( \int \vec{V}_1 dt .\nabla\biggr) \vec{V}_1 \Bigr \rangle 
\end{equation}

These time-averaged second-order variables define the Gor’kov acoustic potential and streaming velocity fields. For a spherical particle much smaller than the acoustic wavelength ($d_p << \lambda_{ac}$), primary acoustic radiation forces can be derived from the gradient of the Gor’kov potential \cite{gorkov_selected_2014} given by,

\begin{equation} \label{eq1}
    U_r = \left[ \frac{1}{2\rho_0 c_0^2}f_m(\tilde{\kappa})\langle p_1^2 \rangle - \frac{3}{4}\rho_0 f_d(\tilde{\rho},\tilde{\delta})\langle \vec{V}_1.\vec{V}_1 \rangle \right]
\end{equation}

as $\vec{F_r} = -V_0 \nabla U_r$.  The terms $f_m(\tilde{\kappa})$ and $f_d(\tilde{\rho},\tilde{\delta})$ are the acoustic monopole and dipole coefficients, respectively. These coefficients are related to the material properties of the particle and the fluid as \cite{bruus_acoustofluidics_2012}, $f_m(\tilde{\kappa}) = 1- \tilde{\kappa}$ and $f_d(\tilde{\rho},\tilde{\delta}) = \bigl( \frac{2(\tilde{\rho}-1)(1-\Gamma(\tilde{\delta}))}{2\tilde{\rho}+1-3\Gamma(\tilde{\delta})} \bigr)$, where, $\tilde{\kappa} = ({\kappa_p}/{\kappa_0})$ , $\tilde{\rho} = ({\rho_p}/{\rho_0})$ , $\tilde{\delta} = ({2\delta}/{d_p})$, and $\Gamma(\tilde{\delta}) = \frac{-3}{2}(1 + i(1+\tilde{\delta}))\tilde{\delta}$. Here, $\rho_p$ and $\rho_0$ represent the densities of the particle and the fluid, respectively, while $\kappa_p$ and $\kappa_0$ represent the compressibilities of the particle and the fluid, respectively. The drag forces acting on cells induced by acoustic streaming can be evaluated as,

\begin{equation}
    \vec{F}_{str} = 3\pi \mu d_p (\langle \vec{V}_{2} \rangle - \vec{V}_p)
\end{equation}

A detailed formulation of the variable decomposition via perturbation theory, including governing equations, boundary conditions, and material properties, is given in \cite{nandy_tunable_2025}.

\subsubsection{ Theoretical scaling of stresses acting on cell membranes}
Various mechanisms of stress generation on cell membranes leading to transient permeabilization are schematically shown in Fig. \ref{fig4}a). Differences in density and compressibility between the intracellular and host fluids cause acoustic scattering and momentum transfer, producing second-order acoustic radiation stress on the cell. The time-averaged acoustic radiation pressure at any point in the fluid scales as:

\begin{equation}
    \langle p_2 \rangle \sim \biggl( {\frac{p_1^2}{4\rho_0 c_0^2} \biggr) \bigl(1+\Gamma_{fl}^2 \bigr) }
\end{equation}

Here, $\rho_0$ and $c_0$ are the host fluid density and sound speed, and $p_1$ and $\Gamma_{fl}$ are the acoustic pressure amplitude and damping factor at the actuation frequency. The radial membrane stress is the difference across the cell, $\sigma_{ar} = \sigma_{ar}^i - \sigma_{ar}^o$, with i and o denoting intra- and extracellular regions. Based on acoustic scattering and transmission, this stress due to radiation forces can be scaled as (Supplementary Information):,

\begin{equation}
    \sigma_{ar} \sim \frac{9 \langle E_{ac} \rangle \tilde{\rho} (\tilde{\rho} - 1)}{(1 + 2\tilde{\rho})^2}
\end{equation}

where, $\tilde{\rho}$ denotes the ratio of density between the cell and the extracellular host fluid. Since acoustic energy density at any frequency of actuation is found to be proportional to the square of the driving voltage, $\langle E_{ac} \rangle \propto V_{p-p}^2$, hence, the acoustic radiation stress acting on the cell membrane is expected to follow the same dependence. The scales of acoustic radiation stress normalized by the square of the peak-to-peak voltage, as a function of frequency, has been presented in Fig. \ref{fig4}(d).

Additionally, viscous dissipation at oscillating fluid–solid interfaces induces time-averaged acoustic streaming, producing streaming-induced stresses that scale as detailed in Supplementary Information:

\begin{equation}
    \sigma_{st} \sim \mu_0 \biggl( \frac{v_{str}}{\delta_v} \biggr)
\end{equation}

Here, $v_{str}$ refers to the typical magnitude of acoustic streaming velocity and $\delta_v = \sqrt{2\mu_0/(\rho_0 \omega)}$ denotes the  hydrodynamic boundary layer thickness formed as a result of viscous dissipation of ultrasonic waves at the oscillating boundaries, namely, at the fluid-cell interface. Analogous to Rayleigh’s parallel-plate streaming \cite{muller_ultrasound-induced_2013}, the magnitude of acoustic streaming velocity $v_{str}$ can be related to the first-order acoustic velocity amplitude $v_1$ as $v_{str} \sim \Psi(\omega) ({v_1^2}/{c_0})$, which yields,

\begin{equation}
    \sigma_{st} \sim \biggl( \frac{1}{4 \sqrt{2}}\biggr) \biggl( \frac{\sqrt{\mu_0} p_1^2 k^2 \Psi(\omega)}{c_0 (\rho_0 \omega)^\frac{3}{2}} \biggr)
\end{equation}

Here, $\Psi(\omega)$ is a frequency-dependent factor relating the first and second-order velocities, and can be derived from fully-coupled numerical simulations. Following similar arguments as described earlier, the streaming-induced stresses on the cell membrane are found to be proportional to the square of the driving peak-to-peak voltage, since, $p_1 \propto V_{p-p}$, and the normalized streaming-induced stress as a function of actuation frequency has been plotted in Fig. \ref{fig4}(d).

Transient changes in acoustic potential drive cells toward potential minima, creating relative motion with the host fluid and resulting in hydrodynamic stresses that scale as:

\begin{equation}
    \sigma_{am} \sim \mu_0 \biggl( \frac{v_{cell} - v_{str}}{\delta_h}\biggr)\sqrt{Re_{cell}}
\end{equation}

where, $v_{cell}$ refers to the acoustophoretic migration velocity of the cell, $\delta_h$ denotes the hydrodynamic boundary layer thickness formed as a result of relative motion between the cell and the host fluid, and $Re_{cell}$ refers to the Reynolds' number associated with the migrating cell, and can be defined as, $Re_{cell} = (2\rho_0 v_{cell} R_{cell}/\mu_0)$. When $v_{cell}>> v_{str}$, the hydrodynamic stress scale simplifies to:

\begin{equation}
    \sigma_{am} \sim \biggl( \frac{F_r}{6\pi R_{cell}^2}\biggr) \sqrt{Re_{cell}}
\end{equation}

Here, $F_r$ is the primary acoustic radiation force in the microcavity. In the overdamped limit, cell velocity scales with $F_r$, so the hydrodynamic stress on the membrane shows a cubic dependence on $V_{p-p}$, as normalized and plotted in Fig. \ref{fig4}(d). In clustered cells, motion of one cell induces hydrodynamic interactions with neighbors, which we scale by considering nearest-neighbor, close-packed arrangements (Supplementary Information):

\begin{equation}
    \sigma_{hi} \sim \frac{13 F_r}{8\pi R_{cell}^2} \sqrt{Re_{cell}}
\end{equation}

The exact hydrodynamic stress depends on cell positions within a cluster. While hydrodynamic forces are long-ranged, introducing cutoffs can yield unphysical results; nevertheless, our scaling captures collective effects by considering nearest-neighbor interactions. 

Due to the dynamic frequency sweep, the total stress on cells accumulates over the cycle and depends on the step size, $\Delta f$ and dwell time, $\tau_{dwell}$. The average membrane stress over one sweep cycle is estimated as:

\begin{equation}
    \bar{\Sigma}_i = \frac{1}{T_{cycle}} \int_{0}^{T_{cycle}} \Sigma_i \,dt\ 
\end{equation}

where, $T_{cycle}$ denotes the total time period of one frequency sweep cycle, while $\Sigma_i$ denotes the instantaneous stress experienced by the cell (either normal or tangential). Following principles derived from linear elasticity \cite{ruderman_fluid_2019}, the equivalent normal stress on the cell membrane can be estimated as,

\begin{equation}
    \bar{\Sigma}_{n,eq} = \frac{1}{2}\sqrt{\bar{\Sigma}_r^2 + 4\bar{\Sigma}_t^2}
\end{equation}

where, $\bar{\Sigma}_r$ and $\bar{\Sigma}_t$ denote the average normal and tangential stresses experienced by the membrane over the entire frequency sweep cycle. The volume incompressibility of cell thus leads to the development of a tension $\sigma_m$ across the membrane under the application of external stresses, which, according to Young Laplace relation can be deduced as \cite{isambert_understanding_1998},

\begin{equation}
    \sigma_m = \bar{\Sigma}_{n,eq} \biggl(\frac{R_{cell}}{2}\biggr)
\end{equation}

\subsubsection{ Scaling of membrane strain energies}
The application of external stress results in deformation of the cell membrane, thus manifesting in the form of strain energy stored in the cell membrane. The seminal work by Helfrich highlights stretching and bending to be the primary modes of deformation of cell membrane. The elastic strain energy of the cell membrane due to stretching can be written as,

\begin{equation}
    E_{m}^{st} = \frac{1}{2}K_A (4\pi R_{cell}^2) \biggl(\frac{\Delta A}{A} \biggr)^2
\end{equation}

where, $K_A$ refers to the area expansion modulus of the cell membrane, and $(\Delta A/A)$ denotes the area strain on the cell membrane. Following the governing equation of membrane deformation as studied in \cite{pak_gating_2015}, the strain energy in the cell membrane due to stretching can be scaled as (details in Supplementary Information),

\begin{equation}
    E_{m}^{st} = \frac{2\pi R_{cell}^2 K_A h_m^4 \sigma_m^2}{[64 \kappa_B + (K_A + 4\sigma_m)h_m^2]^2}
\end{equation}

where, $\sigma_m$ refers to the membrane tension acting on the cell membrane, $\kappa_B$ refers to the elastic bending modulus of the cell membrane and $h_m$ is the thickness of the cell membrane. The bending strain energy on the cell membrane can be obtained using \cite{brown_elastic_2008},

\begin{equation}
    E_{m}^{bnd} = \frac{1}{2}\iint_A \kappa_B \biggl( \frac{1}{R_1} + \frac{1}{R_2} - C_0 \biggr)^2 dA
\end{equation}

where, $R_1$ and $R_2$ are the principal radii of curvature and $C_0$ denotes the spontaneous curvature, while the integral is performed over the entire surface of the cell membrane. Scaling the bending energy with physically relevant parameters, as studied in  \cite{pak_gating_2015} and neglecting the effect of spontaneous curvature, we get (details in Supplementary Information),

\begin{equation}
     E_{m}^{bnd} = \frac{128\pi R_{cell}^2 \kappa_B h_m^2 \sigma_m^2}{[64 \kappa_B + (K_A + 4\sigma_m)h_m^2]^2}
\end{equation}

The total strain energy in the cell membrane can be considered as the summation of contributions due to stretching and bending, $E_{m} = E_{m}^{st} + E_{m}^{bnd}$.

Theoretical scaling of membrane strain energies reveal the interplay between three parameters, namely, the bending modulus, $\kappa_B$, the area expansion modulus, $K_A$, and the equivalent membrane tension developed as a consequence of external field-mediated stresses, $\sigma_m$. 

\subsubsection{Pore formation dynamics}

Stable hydrophilic pore formation on the cell membrane is governed by energy barriers. The work done in forming a pore of radius r results from competition between edge energy and membrane tension \cite{akimov_pore_2017}:

\begin{equation}\label{eq6}
    \Delta W_{pore}(r) = U_0 + \beta \biggl( \frac{r^*}{r} \biggr)^4+ 2\pi\gamma r - \pi r^2 \sigma_{m}
\end{equation}

where, $U_0$ is associated with a nucleation free energy that contributes to the activation energy of pore formation and is found to be independent of the lateral tension \cite{karal_communication_2015}, $\beta$ refers to a steric repulsion energy scale and $r^*$ denotes the metastable pore radius indicating transition from hydrophobic defect to a stable hydrophilic pore size \cite{akimov_pore_2017-1, akimov_pore_2017}, $\gamma$ refers to the edge tension at the periphery of the pore and $\sigma_{m}$ refers to the effective membrane tension. Under a stressed state, there is a reorientation of the lipids, which results in the transition of a hydrophobic \textit{defect} to a hydrophilic pore, which is usually termed as pore inversion. 

 At the initial stage of its lifetime, a pore nucleates at a radius $r_i = (\gamma/\sigma_{m})$, with the characteristic time scale for pore nucleation given by $\tau_{nuc} = (\eta h_m/2\gamma)$ \cite{sandre_dynamics_1999}. Following pore nucleation, subsequent stage of pore growth is governed by the equation,

\begin{equation}
    ln \biggl(\frac{r}{r_i} \biggr) - \frac{1}{2} ln \biggl( \frac{r_c^2 - r^2}{r_c^2 - r_i^2} \biggr) = \biggl( \frac{t}{\tau_{grow}} \biggr)
\end{equation}

with the characteristic time scale for pore growth given by, $\tau_{grow} = (\eta h_m/\sigma_{m})$. In the preceding expressions, $\eta$ refers to the viscosity of the lipid membrane, $h_m$ refers to the thickness of the cell membrane, usually taken as the thickness of the lipid bilayer, and $r_c$ represents the critical radius of pore which results in complete relaxation of tension on the membrane through the formation of pore and release of energy. This critical radius of pore formation can be estimated from the areal strain of the membrane, accounting for the changes in area arising out of thermal fluctuations as well as the externally applied membrane tension \cite{sandre_dynamics_1999},

\begin{equation}
    \biggl( \frac{\pi r_c^2}{A} \biggr) = \biggl( \frac{k_B T}{8\pi \kappa_B}  \biggr) ln\biggl( 1 + \frac{1}{24\pi} \biggl( \frac{\sigma_{m} A}{\kappa _B}\biggr) \biggr) + \biggl( \frac{\sigma_{m}}{K_A}\biggr)
\end{equation}

Here, $\kappa_B$ and $K_A$ refer to the elastic bending modulus and the membrane compressibility modulus, respectively, $k_B^{th}$ is the Boltzmann constant and $T$ refers to the absolute temperature; the constant $(1/24\pi)$ arises for a spherically closed membrane, which is relevant for a biological cell under consideration. The evolution of membrane tension during the pore growth stage is usually described using, $\sigma(r) = \sigma_m (1-(r/r_c)^2)$. This is succeeded by the second stage of lifetime, where the pore grows to reach its maximum radius, $r_m = r_c -(r_i/2)$. Since the initial nucleation radius is found to be much smaller compared to the critical radius of pore \cite{karatekin_cascades_2003}, $r_m \approx r_c$. Henceforth, this maximum radius $r_m$ is considered to be the effective radius of pore formation for the purpose of this study in order to estimate the enhanced diffusivity through the membrane. . In the subsequent stage, leak-out of internal fluid is negligible if $\tau_L>> \tau_{gr}$, consistent with observed cell volume preservation and viability under ultrasound, due to active cellular responses involving $Ca^{2+}$ influx \cite{mcneil_emergency_2005, zhou_effects_2008} and active cytoskeletal modeling \cite{andrews_damage_2014,dias_plasma_2021, jia_spatiotemporal_2022}, directed towards restoration of cellular homeostasis; these mechanisms essentially prevent leakage of components of internal compartments of cell into the extracellular fluid.  This corroborates well with our experimental observations where we do not observe any significant shrinkage in the cell volume and/or loss of cellular viability upon exposure to ultrasonic waves. The fourth stage of pore lifetime is related to the spontaneous closure of the pore upon removal of external stress. Pore closure is usually driven at a constant velocity, $V_{closure} = -\gamma/(2 h_m)$, which results from the balance of line tension energy driving the pore closure and viscous dissipation within the lipid membrane \cite{karatekin_cascades_2003}. The associated time scale of pore closure is given by, $\tau_{closure} \approx (2\eta h_m/\gamma)(r_c -r_i)$. typically much shorter than the interval between actuation cycles, allowing each frequency sweep to generate transient pores without membrane rupture.

\subsubsection{Estimation of diffusivity through permeabilized cell membrane}

Effective diffusivity through the permeabilized membrane is estimated using hindered transport theory \cite{deen_hindered_1987, g_davidson_hydrodynamic_1988}, which accounts for hydrodynamic drag in confined geometries. Considering purely steric interactions, the macroscopically averaged flux can be expressed as:

\begin{equation}
    \bigl[J_p] = \biggl( \frac{6\pi \bigl[\phi_p] }{K_t}\biggr) \biggl( \frac{D_{\infty}}{h_m}\biggr) \Delta C_p
\end{equation}

\begin{equation}
    \bigl[\phi_p] = \bigl( 1-\lambda_p)^2
\end{equation}

where, $\bigl[\phi_p] $ refers to the partitioning coefficient, $\lambda_p = (r_{sol}/r_{pore})$ is the ratio of the hydrodynamic radius of solute molecule to the characteristic size of pore formed on the membrane surface, $D_\infty = (k_B T)/(6\pi \mu r_p)$ is the diffusion coefficient in unbounded fluid medium estimated using Stokes-Einstein relation and $\Delta C_p$ refers to the difference in concentration of solute across the membrane. The term $K_t$ is associated with a hydrodynamic drag coefficient that accounts for the enhanced drag experienced by the solute molecules while traversing through the confined geometry of the membrane pore. The hydrodynamic coefficient $K_t$ has been established from the works of \cite{bungay_motion_1973} and its expanded expression can be found in \cite{deen_hindered_1987}.

Once the diffusivity through a single pore is determined, the effective diffusivity through the porous membrane is obtained by accounting for the fraction of the surface covered by pores. Estimating permeability for finite-thickness membranes is non-trivial and was recently addressed in \cite{skvortsov_permeability_2023}. For arbitrary pore distributions, the effective diffusivity is given by:

\begin{equation}
    D_{eff} = \biggl( \frac{2\zeta_{pore}D_{ch}D_\infty h_m}{2D_\infty h_m + \pi r_m M(\zeta_{pore}) D_{ch}}\biggr) 
\end{equation}

where, $\zeta_{pore} = (A_{pore}/A_m)$ is the ratio of effective area occupied by pores to the membrane surface area, $D_{ch}$ refers to the single-channel diffusivity through the pore as estimated using hindered diffusion theory. According to the aforementioned proposed scaling analysis, $A_{pore} = N_m(\pi r_m^2)$. The functional dependence is specified as, $M_{pore} = (1-\zeta_{pore})^2 /(1 + A_1 \sqrt{\zeta_{pore}} - B_1 \zeta_{pore}^2)$, where the empirical constants $A_1$ and $B_1$ have been specified as $0.34$ and $-0.58$, respectively, for an arbitrary distribution of pores on the membrane surface. This framework enables evaluation of enhanced diffusivity under ultrasonic-induced permeabilization. The corresponding timescale for transport across the permeabilized membrane is estimated as (Supplementary Information): 

\begin{equation}
    \tau_{df}^{ac} = \frac{R_{cell}^2}{3\zeta_{pore}(\sigma)D_{eff}(\sigma)}
\end{equation}

where, $D_{eff}$ corresponds to the enhanced diffusivity coefficient for the particular biomolecular species as estimated in the preceding section, and  $\zeta_{pore}$ represents the area occupied by the pores as a fraction of the total membrane surface area; both these parameters have implicit dependence on the membrane tension, $\sigma$. This timescale assumes that, after membrane permeation, the biomolecule uniformly occupies the cellular volume, neglecting cytosolic transport and nuclear envelope permeation for simplicity.

\section*{Acknowledgements}
A. K. S. expresses gratitude to the Department of Science \& Technology (DST), Government of India, for the financial support through the Swarnajayanti Fellowship Award under Grant No. DST/SJF/ETA-03/2017-18, and acknowledges the support provided by the Ministry of Education, Government of India, under IoE Phase II.

\section*{Author contributions}
S. N. designed the research framework, performed experiments and numerical simulations, carried out data analysis, and wrote the paper. A. K. S. devised the core idea of the research, designed the research framework, supervised the research, gave key suggestions to improve the research, and wrote the paper. M. M. cultured the biological cells and helped in the experiments.

\section*{Competing Interests}
The authors declare no competing interests.
\\

\hrule
\newpage

\section*{Supplementary Information for\\
    \textbf{ "Programmable ultrasonic fields enhance intracellular delivery in cell clusters"\\
    }
}
\begin{center}
    Subhas Nandy, Monica Manohar, Ashis K Sen* 
\end{center}
\hrule

\setcounter{equation}{0}
\setcounter{section}{0}
\setcounter{figure}{0}
\renewcommand{\figurename}{Fig.S}

\section{Supplementary Note 1:}
\subsection{Control experiments with cells}
In order to compare the ultrasonic-mediated transmembrane transport of biomolecular cargos against untreated cells, control experiments are performed with HeLa cells; the results have been presented in Fig.S 2. These cells are trypsinized from the cell culture flask, washed and resuspended in fresh cell culture medium. Different biomolecular cargos are then added into the medium. The cells are suspended in a petri dish mounted over a confocal microscope setup and imaging is performed periodically over 6 h. Closer to the initial time period, fluorescence emission by propidium iodide (PI) and doxorubicin (DOX) remains below detectable thresholds and cannot be quantified. The fluorescence emission intensities increase marginally after a period of 30 mins under TRITC imaging filter; this highlights very slow permeation of PI and DOX across the cell membrane. Similarly, no fluorescence intensity changes are detected in FITC imaging filter, thus signifying very slow efflux of calcein across the membrane. Control experiments thus reveal the large time scales associated with passive diffusion, which proves to be an impediment for rapid drug response studies.

\subsection{Acoustic trapping within the microfluidic cavity}
The proposed acoustofluidic device offers capabilities of simultaneous ultrasound-mediated trapping of biological cells and intracellular delivery. In order to illustrate the trapping capacity of the device, cell suspensions of different initial cell densities were prepared and injected into the device using high-precision syringe pumps, followed by aggregation induced by progressive trapping of incoming cells. Fig.S 3 demonstrates the acoustic trapping of cellular aggregates of increasing initial cell seeding densities. This experimentally validates the ultrasonic trapping capabilities and highlights the potential for production of large cellular aggregates and simultaneous exposure of large number of cells to drug-loaded microenvironments. 

\subsection{Intracellular delivery of different biomolecular agents}
In order to experimentally validate the non-specific nature of intracellular delivery of different biomolecular agents, additional experiments were performed with different combinations of compounds. Imaging is performed under confocal microscopy to enhance visualization and colocalization of fluorescence emission spots within the volume of the cellular aggregate. Fig.S 4(a) illustrates the concomitant influx and efflux of DAPI and Calcein into and out of the cells with increasing number of frequency sweep cycles, respectively. Another combination of staining fluorophore-based reagents is used, namely, Hoescht-33342 and actin staining dye, as presented in Fig.S 4(b). Initially, Hoescht-33342 is added into the fresh cell suspension medium and incubated for a period of 15 minutes. This is followed by addition of actin staining dye and immediate injection into the microfluidic device, without any period of incubation. With increasing number of frequency sweep cycles, the actin staining dye, as observed under red fluorescence emission, is observed to progressively get transported into the intracellular volume, thus establishing the non-specific nature of biomolecular transport of the proposed methodology.

\subsection{Intracellular delivery corresponding to different cell lines}

In order to understand the influence of frequency sweeps on ultrasound-mediated intracellular delivery across different cell lines, additional experiments were performed with cervical cancer cells, HeLa and breast cancer cells, MCF-7. Experimental snapshots of fluorescence intensity evolution representative of the two cell lines have been presented in Fig.S 6. The uptake and efflux trends exhibited by the two cell lines under investigation deserves some discussion. The uptake of DOX by MCF-7 cells is observed to be faster as compared to HeLa cells. Transmembrane transport and subsequent intercalation of DOX with nucleic acids present within the intracellular space is governed by a number of factors, including fluidity of cell membrane and other intrinsic factors maintaining the cell microenvironment. Of particular importance in this regard, is the membrane potential \cite{yang_membrane_2013}. Prior studies have established that the infliction of a wound on the cell membrane by imposition of external stresses triggers the activation of mechanosensitive (MS) channels and transport of ions across the membrane. In certain cell lines, the triggering of these MS channels results in a simultaneous Ca\textsuperscript{2+} influx and K\textsuperscript{+} efflux; when the K\textsuperscript{+} efflux overcompensates for the Ca\textsuperscript{2+} influx, the cell membrane tends to become hyperpolarized \cite{juffermans_low-intensity_2008}. Similar observations were made by \cite{tran_effect_2007}, wherein, cell membrane hyperpolarization was recorded for breast cancer cells. An enhanced transmembrane potential arising out of hyperpolarization has been observed to enhance the penetration of DOX across the cell membrane \cite{aminipour_passive_2020}. It has also been hypothesized that the hyperpolarization of cell membrane facilitates the phenomenon of endocytosis, thus enabling the uptake of large macromolecules and subsequent intracellular transport and DNA intercalation \cite{juffermans_low-intensity_2008,zupancic_differential_2002}. However, the situation is different for HeLa cells. It has been established that HeLa cells are deficient in voltage-gated ion channels. Hence, disruption of cell membrane for HeLa cells leads to transient or permanent depolarization (depending on external stresses imposed) \cite{qin_sonoporation-induced_2014,wen_ultrasound_2023}. This leads to the reduction in transmembrane potential, eventually hindering the transport of DOX across the cell membrane. We hypothesize that these differences in transmembrane potentials, upon ultrasonic-mediated membrane disruption, potentially governs the faster uptake trends observed for DOX in MCF-7 cells as compared to HeLa cells. It should however be noted that our experimental observations primarily focus on the uptake trends during the frequency sweep operation cycles, and does not attempt to make any comparison of the retention of DOX by the cells post ultrasonic treatment. It has been established in numerous studies that, inherently, MCF-7 cells exhibit higher resistance to the retention of DOX in the intracellular volume as compared to HeLa cells \cite{aminipour_passive_2020}; this is primarily attributed to the higher expressions of different efflux transporters in MCF-7 cells, namely P-glycoprotein (P-gp), which upon removal of externally imposed membrane tension, actively transports internalized doxorubicin molecules back to the extracellular fluid across the cell membrane \cite{adkins_p-glycoprotein_2013,pasquier_p-glycoprotein-activity_2013,robey_revisiting_2018}. However, there have been a growing number of experimental evidences which seem to emphasize that ultrasound exposure can significantly inhibit the expression of P-gp in cancer cells \cite{wu_ultrasound_2011, huang_effects_2018,zhang_low_2013,du_impact_2022,aryal_effects_2017}, thus enhancing the accumulation of chemotherapeutic agents within the cellular volume and eventual killing of cancerous cells through drug-induced cytotoxicity. Comparative study of the efficacy of various drug treatments across different cancer cell lines remains the scope of future study.

\section{Supplementary Note 2:}
\subsection{Scaling of Acoustic Radiation Stress}

Considering the two-dimensional acoustic pressure distribution within the microcavity volume to be given by an analytical relationship of the form,

\begin{equation}
    p_i(x,y) = A_i cos(k_{x,i}x + k_{y,i}y + \phi_i)
\end{equation}

where, $A_i$,$k_{x,i}$,$k_{y,i}$ and $\phi_i$ refer to the acoustic pressure amplitude, wavevector components along X and Y-directions and phase of the ultrasonic wave, corresponding to the $i$-th frequency during a frequency sweep cycle. Converting the analytical expression to polar coordinates and performing a Jacobi-Anger expansion results in the incident pressure field given by,

\begin{equation}
    p_{in}(r,\theta) = A_i \sum_{n=0}^{\infty} J_n(k_0r)cos(n(\theta - \alpha_0) + \phi_i + (n\pi/2))
\end{equation}

where, $k_0 = (\omega/c_0)$ refers to the wavenumber in the fluid medium, $\alpha_0 = tan^{-1}(k_{y,i}/k_{x,i})$ and $J_n$ refers to the Bessel function of order $n$. Considering the scattering from a cell approximated as a spherical body, the scattered and transmitted acoustic pressure field distributions can be written as \cite{doinikov_acoustic_1997},

\begin{equation}
    p_{sc}(r,\theta) = A_i \sum_{n=0}^{\infty} s_n H_n^{1}(k_0r)cos(n(\theta - \alpha_0) + \phi_i + (n\pi/2))
\end{equation}

\begin{equation}
    p_{tr}(r,\theta) = A_i \sum_{n=0}^{\infty} t_n J_n(k_cr)cos(n(\theta - \alpha_c) + \phi_i + (n\pi/2))
\end{equation}

where, $s_n$ and $t_n$ refer to the scattered and transmitted coefficients, corresponding to the $n-$th wave component, respectively. The terms $k_c$ and $\alpha_c$ denote the wavenumber and wave propagation direction inside the spherical cell, respectively. Equating the first-order acoustic pressure and velocities across the interface $r=a$ results in a linear system of equations for the scattering and transmission coefficients, which are given by,

\begin{equation}
    s_n = \frac{k_0J_n(k_ca)J_n^{'}(k_0a) - k_c J_n^{'}(k_c a)J_n(k_0 a)}{k_c H_n^{1}(k_0 a) J_n^{'}(k_i a) - k_0 J_n(k_c a) H_n^{1'}(k_0 a) }
\end{equation}

\begin{equation}
    t_n = \frac{k_0 J_n(k_0a) H_n^{1'}(k_0 a) - k_0 J_n^{'}(k_0 a)H_n^{1}(k_0 a)}{k_0 J_n(k_c a) H_n^{1'}(k_0 a) - k_c J_n^{'}(k_c a) H_n^{1}(k_0 a) }
\end{equation}

Here, the prime $'$ denote differentiation with respect to the quantity encloses within parentheses . The first-order velocity fields inside ($c$) and outside ($0$) the spherical body are given by,

\begin{equation}
    \vec{V}_0 = \frac{-i}{\rho_0 \omega}\nabla(p_{in} + p_{sc})
\end{equation}

\begin{equation}
    \vec{V}_c = \frac{-i}{\rho_c \omega}\nabla(p_{tr})
\end{equation}

which can be calculated once the scattering and transmission coefficients are estimated. Knowing the first-order acoustic pressure and velocities across the surface of the cell-fluid interface, the acoustic radiation pressure can be calculated as, $\sigma_{rr, r=a} = \sigma_{rr}^{0} - \sigma_{rr}^{c}$, where, $\sigma_{rr}$ refers to the radial component of the acoustic radiation stress given by,

\begin{equation}
    \sigma_{rr}^{k} = \frac{1}{4}\biggl(  \rho_k \langle \vec{V}_{1,k} . \vec{V}_{1,k} \rangle - \kappa_k |p_{1,k}|^2 \biggr) - \frac{1}{2}\rho_k |V_{1,k}^r|^2
\end{equation}

where, $k=0,c$ denote the inner and outer medium respectively, $\rho_k$ and $\kappa_k$ denote the density and compressibility of the $k$-th medium, while, $V_{1,k}^r$ denotes the radial component of the first-order velocity at the cell-fluid interface. Substituting the expressions and performing a scaling analysis leads to a theoretical scale for the acoustic radiation stress given by,

\begin{equation}
    \sigma_{ar} = \frac{9 \langle E_{ac} \rangle \tilde{\rho} (\tilde{\rho} - 1)}{(1 + 2\tilde{\rho})^2}
\end{equation}

\subsection{Scaling of stress due to acoustic streaming}

Due to the formation of a viscous boundary layer at the cell-fluid interface as a result of ultrasonic irradiation and scattering, viscous stresses are exerted on the cell membrane, which can be estimated as \cite{sadhal_acoustofluidics_2012},

\begin{equation}
    \sigma_{sf} = \mu_0 \frac{V_{str}(\omega)}{\delta_v(\omega)} \sim \mu_0 \frac{\psi(\omega) V_1^2(\omega)}{c_0 \delta_v(\omega)}
\end{equation}

where, $V_{str}$ refers to the magnitude of acoustic streaming velocity and $\delta_v = \sqrt{2\mu_0/\rho_0 \omega}$ refers to the viscous boundary layer thickness, at a given angular frequency $\omega$. Drawing analogy to Lord Rayleigh's scaling analysis for acoustic streaming between parallel plates, the second-order streaming velocity can be related to the first-order acoustic velocity in the fluid medium as, $V_{str} \sim \psi(\omega) V_1^2/c_0$, where, $\psi(\omega)$ is a geometry and frequency dependent factor which can be estimated from numerical simulations. Following the analytical expression derived in preceding section, the first-order acoustic velocity can be related to the first-order acoustic pressure amplitude, and the corresponding scales are given by, $V_1 \sim {p_1}/{\rho_0 \omega \lambda}$, which yields the scaling relationship for acoustic streaming-induced stress as,

\begin{equation}
    \sigma_{sf} \sim \frac{p_1^2 k^2 \sqrt{\mu_0} \psi(\omega)}{4\sqrt{2} \pi^2(\rho_0 \omega)^{\frac{3}{2}} c_0}
\end{equation}

\subsection{Scaling of hydrodynamic interaction stress}

For two equally-sized spherical particles $d_p = 2a$ interacting hydrodynamically with each other subjected to external forces, the mobility tensor $\underline{\underline{\mu_{ij}}}$ between non-overlapping spheres can be given by the Rotne-Prager-Yamakawa (RPY) tensor as \cite{ichiki_many-body_2001},

\begin{equation}
    \underline{\underline{\mu_{ij}}} = \frac{1}{8\pi \mu_0 |r_{ij}|}\biggl[\biggl( \underline{\underline{I}} + \frac{\underline{r_{ij}} \times \underline{r_{ij}}}{|r_{ij}|^2}\biggr) + \frac{2a^2}{|r_{ij}|^2}\biggl( \frac{\underline{\underline{I}}}{3} - \frac{\underline{r_{ij}} \times \underline{r_{ij}}}{|r_{ij}|^2}  \biggr) \biggr]
\end{equation}

where, $\underline{r_{ij}}$ is the vector directed from cell $i$ to cell $j$, and $|r_{ij}|$ denotes the absolute value of the distance separating the cells and $\underline{\underline{I}}$ represents the identity tensor. The corresponding velocities of the cells can be obtained as, $\underline{V_{ij}} = \underline{\underline{\mu_{ij}}}.\underline{F_j}$, where, $\underline{F_j}$ denotes the force acting on particle $j$ at the current instant. Considering only translational components, the velocities of the cells can be written as,

\begin{equation}
    V_{ij,x} = \frac{1}{8\pi \mu_0 |r_{ij}|} \biggl[ \biggl( 1 + \frac{r_x^2}{|r_{ij}|^2} + \frac{2a^2}{3|r_{ij}|^2} - \frac{2a^2r_x^2}{|r_{ij}|^4} \biggr)F_{j,x} + \biggl( \frac{r_x r_y}{|r_{ij}|^2} -  \frac{2a^2 r_x r_y}{r_{ij}^4}\biggr)F_{j,y} \biggr]
\end{equation}

where, $r_x$ and $r_y$ are the X- and Y-components of the direction vector from cell $i$ to cell $j$. Performing a scaling analysis leads to $r_x \sim r_y$ and $F_{j,x} \sim F_{j,y}$, which results in the hydrodynamic interaction velocity between cell $i$ and $j$ given by,

\begin{equation}
    \underline{V_{ij}} = \frac{1}{8\pi \mu_0 |r_{ij}|} \biggl( 3 - \frac{10a^2}{3 |r_{ij}|^2}\biggr) F_j
\end{equation}

Owing to the $1/|r|$ dependence of the hydrodynamic interaction velocities, we assume that the stronger interactions would primarily occur between cells that are touching each other and are subjected to ultrasonic forces. This leads to scaling the interparticle distance as $|r_{ij}| \sim 2a$, and the corresponding interaction velocity between two cells is given by, $V_{ij} \sim (13 F_j/96\pi\mu_0 a)$. The corresponding hydrodynamic interaction stress is given by,

\begin{equation}
    \sigma_{hi} = \mu_0 \frac{V_{int,hyd}}{\delta_h} \sim \frac{13 F_{j} \sqrt{Re_{cell}}}{96\pi a^2}
\end{equation}

The coordination number of the packing configuration determines the number of cells in contact with a given cell. Assuming a close-packed configuration, the overall hydrodynamic interaction stress on each cell can be scaled as, $\sigma_{hi} \sim (13F_{rad} \sqrt{Re_{cell}}/8\pi \mu_0 a^2)$.

\subsection{Scaling of energies in cell membrane}

The governing equation of deformation of cell membrane under the influence of external field-mediated stresses, neglecting the influence of spontaneous curvature, is given by \cite{pak_gating_2015},

\begin{equation}
    \nabla^4u + \frac{K_A}{\kappa_B a^2}u - \frac{\nabla.(\sigma_m \nabla u)}{\kappa_B} + \frac{\sigma_m}{\kappa_B a} = 0
\end{equation}

where, $u$ refers to the membrane deformation, $a = h_m/2$ is half of the membrane thickness parameter and $K_A$ and $\kappa_B$ refer to the area expansion modulus and bending modulus of cell membrane, respectively. In order to perform a scaling analysis, we assume that a homogeneous, isotropic tension is developed across the cell membrane as a result of ultrasonic and hydrodynamic-mediated stresses, and hence, the parameter $\sigma_m$ is considered to be a constant.

Expanding the deformation of the cell membrane in spherical harmonics in 2D, we have,

\begin{equation}
    u = \sum_{n=2}^{\infty} \biggl( a_n cos(n \theta) + b_n sin(n \theta) \biggr)
\end{equation}

where, $\sqrt{a_n^2 + b_n^2}$ denotes the amplitude of the $n$-th deformation mode. Considering $n=2$ to be the dominant mode of deformation, upon substitution into the governing equation, we obtain,

\begin{equation}
    u \sim \frac{\sigma_m h_m^3}{128 \kappa_B + (2K_A + 8\sigma_m)h_m^2}
\end{equation}

The overall strain energies stored in the cell membrane can thus be scaled as,

\begin{equation}
    E_b = \frac{1}{2}\kappa_B(\nabla^2u)^2 (4\pi R_0^2) \sim \frac{128\pi R_0^2 h_m^2\kappa_B \sigma_m^2}{\bigl(64\kappa_B + (K_A + 4\sigma_m)h_m^2 \bigr)^2}
\end{equation}

\begin{equation}
    E_s = \frac{1}{2}\kappa_B(\nabla^2u)^2 (4\pi R_0^2) \sim \frac{2\pi R_0^2 h_m^4 K_A \sigma_m^2}{\bigl(64\kappa_B + (K_A + 4\sigma_m)h_m^2 \bigr)^2}
\end{equation}

Substituting numerical values for material properties associated with cell membranes of cancer cells, and typical values of membrane tension, yields membrane energy scales that are significantly higher compared to thermal fluctuation energy scales, $E_{thermal} \sim k_BT$.

\section{Supplementary Note 3:}

\subsection{Effective number and area of pores formed under ultrasonic irradiation}

The formation of pores on cell membrane is inherently stochastic in nature. In the absence of external stresses, pore formation is driven purely by thermal fluctuations. Under the application of external stresses, the pore nucleation energy barrier is reduced, thus accentuating the process of hydrophobic-to-hydrophilic transition and subsequent formation of a stable pore across the cell membrane. 

Though pore formation is probabilistic, a scaling analysis can be performed to estimate the maximum number of pores formed on the surface of the cell membrane in response to an external stress. A nucleated pore, under the application of an external stress, keeps growing till a maximum radius is achieved. The growth of pores eventually leads to the relaxation of membrane tension \cite{sandre_dynamics_1999}. This is governed by the areal strain acting on the cell membrane and is given by, $\sigma = \sigma_m - K_A (\Delta A/A_0)$, where, $\Delta A$ denotes the change in area of the cell membrane in response to external stress. The effective force acting at the periphery of the pore can be obtained as, $f_{pore} = \sigma (2\pi r_m) - \gamma$. Assuming that all pores formed on the cell membrane grow to a maximum radius $r_m$ to achieve the condition of dynamic equilibrium, 

\begin{equation}
    \sigma_m - \frac{K_A(\pi r_m^2)N_m}{4\pi R_{cell}^2} = 0
\end{equation}

where, $N_m$ denotes the maximum number of pores formed across the cell membrane. Rearrangement of the aforementioned equation yields,

\begin{equation}
    N_m \sim  \frac{4 \sigma_m R_{cell}^2}{K_A r_m^2}
\end{equation}

This approach can be used to estimate the maximum number of pores formed across the cell membrane, assuming that all pores are formed and grow simultaneously under the application of external stress.

\subsection{Effective time scale for diffusion through permeated cell membrane}

Owing the accentuated formation of pores and subsequent enhancement of diffusivity, the diffusive flux for any biomolecular species across the permeated cell membrane can be estimated as,

\begin{equation}
    j_c \sim \biggl( \frac{D_{eff} A_{eff}}{R_{cell}} \biggr) \Delta c
\end{equation}

where, $A_{eff}$ and $D_{eff}$ refer to the increased area of permeation and enhanced diffusivity coefficient, respectively, while $\Delta c$ denotes the differences in the concentration of the molecular species under investigation across the cell membrane. The transport of species across the permeated membrane is then governed by the first-order kinetics given by,

\begin{equation}
    \frac{4\pi R_{cell}^3}{3} \biggl( \frac{d \Delta c}{dt}\biggr) = \pm \biggl( \frac{D_{eff} A_{eff}}{R_{cell}} \biggr) \Delta c
\end{equation}

where, the $\pm$ symbol indicates influx and efflux across the cell membrane, respectively. Solving the equation yields, $\Delta c(t) = \Delta c_0 .e^{-t/ \tau_{df}^{ac}}$, where, $\Delta c_0$ denotes the initial concentration difference of the biomolecular species across the cell membrane, while, $\tau_{df}^{ac}$ denotes the time scale of transport associated with the enhanced diffusion across the permeated membrane. This time scale can thus be explicitly written as,

\begin{equation}
    \tau_{df}^{ac} \sim \biggl(\frac{4\pi R_{cell}^4}{3 D_{eff} A_{eff}} \biggr)
\end{equation}

\setlength{\tabcolsep}{10pt}
\renewcommand{\arraystretch}{1.2}
\begin{table}[t]

    \centering
    \begin{tabular}{c|c|c|c} 
         \textbf{Compound}&  \textbf{Hydrodynamic diameter}&  \textbf{Unit}\\ 
         \cmidrule{1-3}
         Calcein \cite{tambutte_calcein_2012}&  0.65&  nm\\ 
         DAPI \cite{petersen_chromatic_2004}&  1.5&  nm\\ 
         Propidium iodide (PI)  \cite{yoon_direct_2016}&  1.5&  nm\\ 
         Doxorubicin (DOX)  \cite{karatasos_self-association_2013}&  2.774&  nm\\ 
         
         \cmidrule{1-3}
    \end{tabular}
    \caption{Table containing properties of biomolecular compounds used in study.}
    \label{table2}
\end{table}

\section{Supplementary Note 4:}

\subsection{Highlights of the current study}
In this study, we experimentally validate the capability of the acoustofluidic device in achieving acoustic trapping of suspended cells, followed by exposure to synergistic ultrasonic and hydrodynamic stresses upon being subjected to dynamic alterations in acoustic potential landscape. Theoretical scaling and numerical computations further enable us to study the permeabilization phenomenon and decipher the underlying governing mechanisms behind such transient perforation strategies. It is imperative to draw distinctions of the proposed methodology that distinguishes it from other intracellular delivery strategies based on ultrasound.

While acoustic streaming-based perforation methods have been well established in literature \cite{kim_acoustofluidic_2022, zhang_hypersonic_2017, guo_controllable_2021, ramesan_acoustically-mediated_2018}, such strategies are found to perform optimally for adhered cells. Though these methods can also be adopted for suspended cells, occlusion of rare subpopulations of cells from the field-of-view often renders such techniques inconvenient for downstream analysis of specific sets of suspended cells. Our proposed methodology, in contrast, enables simultaneous trapping and manipulation of suspended cells, thus retaining the cells under investigation within the field-of-view and assisting in downstream analysis.

Ultrasound contrast agent (UCA)-based poration strategies have largely been employed for permeabilization of cell membranes. However, such methods exhibit drawbacks associated with permanent membrane damage owing to violent fluctuations, non-repeatability due to improper control over generated microbubble sizes and complex preparation protocols in terms of coating microbubble surfaces with specific agents to enable adherence to cells \cite{van_elburg_dependence_2023, baig_influence_2024, tu_ultrasound-mediated_2022, kudo_sonoporation_2009}. In addition, such mechanisms demand control over arrays of microbubbles for effective intracellular delivery to large groups of cells \cite{meng_sonoporation_2019}. Our proposed methodology overcomes these drawbacks by circumventing the need of UCA-based sonoporation. This study highlights the simultaneous treatment of acoustically aggregated cells, thus enhancing throughput for downstream drug efficacy analysis. The phenomenon is based on ultrasonic radiation forces and streaming velocity distributions, which are ubiquitous in nature for any ultrasonic phenomenon in microfluidics, and thus precludes the need for additional microbubble-based sonoporation. In addition, the cellular viability of acoustically treated groups has also been validated experimentally, both under short and long-time scales, thus establishing biocompatibility of the phenomenon.

While some recent studies have also incorporated the effects of acoustically induced temperature rise effects to enhance permeabilization \cite{liu_acoustothermal_2024}, such methodologies work optimally under a very narrow range of operating conditions (specifically, temperature gradients) and deviations from these conditions is likely to result in aggravated cell damage. Our experimental results, in contrast, demonstrate the temperature limits of the acoustofluidic device and enable operation under conditions that prevent any significant temperature rise, while ensuring intracellular delivery over reduced time scales as compared to passive diffusion.

The proposed methodology offers exciting potentials of massive upscaling in terms of convenient fabrication using standard photolithography techniques as well as integration with upstream detection systems \cite{gaikwad_optomicrofluidic_2021} for suitable separation of cancerous cells from healthy cells, thus assisting in localized acoustic trapping and drug efficacy studies.\\

\section{Supplementary Movies}

\textbf{Movie S1:} Acoustic trapping of HeLa cells in microfluidic cavity at the operating resonance frequency ($f_{res}$=0.950 MHz, $P_{in}$=2.0 W). Cells are injected into the device till a homogeneous distribution is achieved in microcavity volume. Ultrasonic excitation leads to trapping of cells along the principal diagonal of the microcavity. Subsequent initiation of background fluid flow leads to incoming cells within the microcavity volume. Strong localized ultrasonic radiation forces lead to acoustic trapping and growth of cellular aggregate for subsequent exposure to dynamic alterations in excitation frequency. This video corresponds to the image sequence presented in Fig. 1(b).\\

\textbf{Movie S2:} Response of polystyrene microparticle aggregate subjected to dynamic changes in actuation frequency. The operating parameters include, $\Delta$f = 10 kHz, $\tau_d$ = 1 second, $P_{in}$ = 2.0 W. Microparticle aggregate undergoes changes in morphology, as well as combined translation and rotation under the influence of alterations in excitation frequency. This video qualitatively illustrates the functioning of the proposed acoustofluidic methodology.\\

\textbf{Movie S3:} Response of an ultrasonically trapped HeLa cell aggregate subjected to dynamic changes in actuation frequency. The operating parameters include, $\Delta$f = 10 kHz, $\tau_d$ = 1 second, $P_{in}$ = 2.0 W. Due to stronger cell-to-cell binding, the cellular aggregate retains its original morphology, while undergoing translation and rotation due to localized forces within the microcavity volume.\\

\textbf{Movie S4:} Experimental video showing the influx of DAPI stain into the intracellular volume as a result of exposure to dynamic frequency excitations. Initial frequency sweep cycles exhibit low fluorescence emission intensities, signifying smaller amounts of DAPI in the intracellular volume. Increasing number of frequency sweep cycles results in transient perforation and enhancement of diffusivity through the cell membrane, as reflected in the higher fluorescence intensities corresponding to 15 frequency sweep cycles. This video corresponds to intracellular influx of DAPI stain, as quantified in Fig. 2(c).\\

\textbf{Movie S5:}  Experimental video showing bidirectional transport of calcein and propidium iodide (PI) across the permeabilized cell membrane upon exposure to dynamic frequency excitations. Each frequency sweep cycle is visualized under FITC imaging filter, corresponding to fluorescence emission by calcein, and is succeeded by a snapshot under TRITC imaging filter, corresponding to fluorescence emission by PI. With increase in frequency sweep cycles, fluorescence emission for calcein pertaining to the cell aggregate volume gradually reduces, signifying efflux of fluorescent calcein; this is accompanied by a concomitant increase in fluorescence exhibited by the host fluid medium. In contrast, fluorescence emission for PI within the intracellular volume monotonically increases, signifying intracellular delivery of otherwise-permeant PI molecules. This video corresponds to the bidirectional transport, as quantified in Fig. 2(c).\\

\textbf{Movie S6:}  Experimentally observed dynamic temperature changes of the acoustofluidic device subjected to a driving power of $P_{in}$ = 3.2 W, recorded using infrared imaging. Ultrasonic actuation period is set as $\tau_{on}$ = 30 seconds, followed by a resting period of $\tau_{off}$ = 90 seconds. The temperature rise of the device is significantly high, and provides a plausible mechanism of permanent cellular damage and loss of viability. In addition, the device does not cool down to its initial temperature over the resting period, leading to persistent increase in operating temperatures in subsequent frequency sweep cycles. This video corresponds to the experimental snapshots and quantification presented in Fig. 3(a).\\

\textbf{Movie S7:} Experimentally observed dynamic temperature changes of the acoustofluidic device subjected to a driving power of $P_{in}$ = 2.0 W, recorded using infrared imaging. Ultrasonic actuation period is set as $\tau_{on}$ = 30 seconds, followed by a resting period of $\tau_{off}$ = 90 seconds. The temperature rise of the device during the actuation period, corresponding to this driving power, remains within acceptable limits and does not adversely impact cellular viability. During resting period, the device cools down back to its initial temperature, thus preventing any accumulation of heat over one complete cycle. This video corresponds to the experimental snapshots and quantification presented in Fig. 3(a).\\

\textbf{Movie S8:} Experimental video of acoustic streaming within the microcavity volume at the experimentally observed resonant frequency of acoustofluidic device. Acoustic streaming is visualized using fluorescent polystyrene microparticles of 1 $\mu$m nominal diameter under TRITC imaging filter. The separatrix for the two streaming vortices is observed to be oriented along the auxiliary diagonal of the microcavity, thus highlighting the non-trivial nature of the acoustic streaming distribution and hence, the streaming-induced stresses on cells. This video corresponds to the theoretical estimation of stresses presented in Fig. 4(b). \\

\begin{figure*}[t]
    \centering
    \includegraphics[width=1\linewidth]{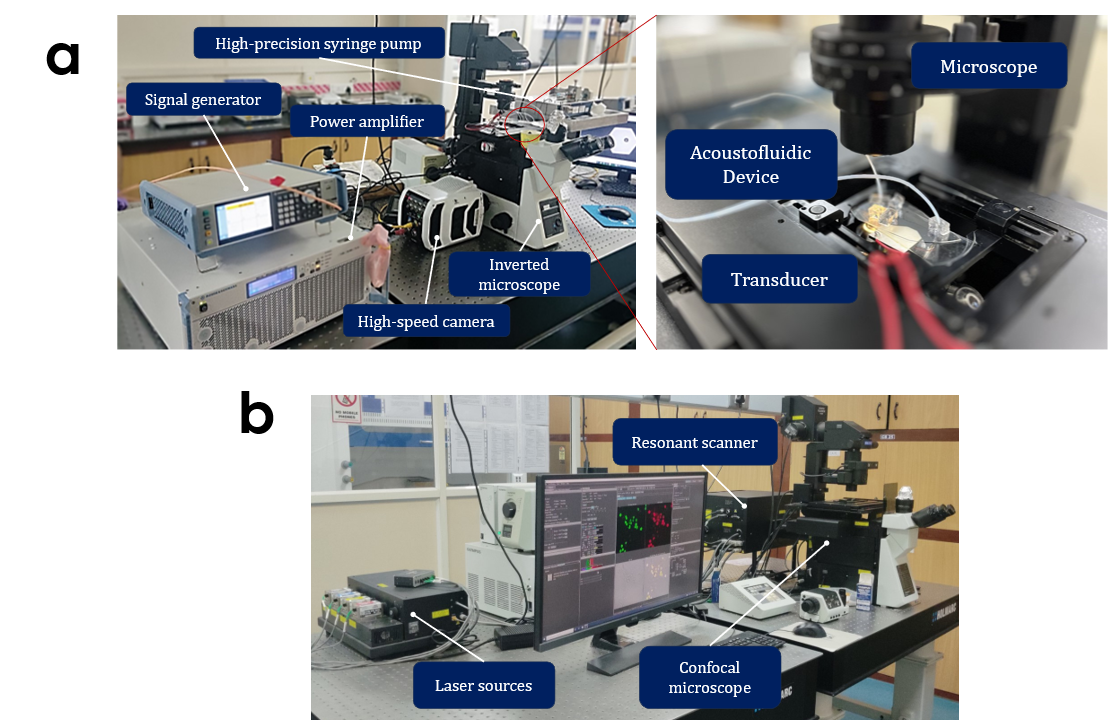}
    \caption{\textbf{Images of experimental setup.} a) Image showing the inverted microscope setup along with high-speed camera used for experiments. The microscope is equipped with a epifluorescence lamp and different fluorescence emission filters that enable visualization of emitted fluorescence intensities from ultrasonically treated cells. The image on the right shows a close-up view of the mounted acoustofluidic device equipped with a piezoelectric actuation source. The device is driven by a signal generator and a power amplifier, b) Image showing the experimental setup of a confocal scanning microscope. The microscope is equipped with laser sources corresponding to different excitation wavelengths, and a resonant scanning system that permits visualization across the cross-section of the sample.}
    \label{fig_S1}
\end{figure*}

\begin{figure*}[t]
    \centering
    \includegraphics[width=1\linewidth]{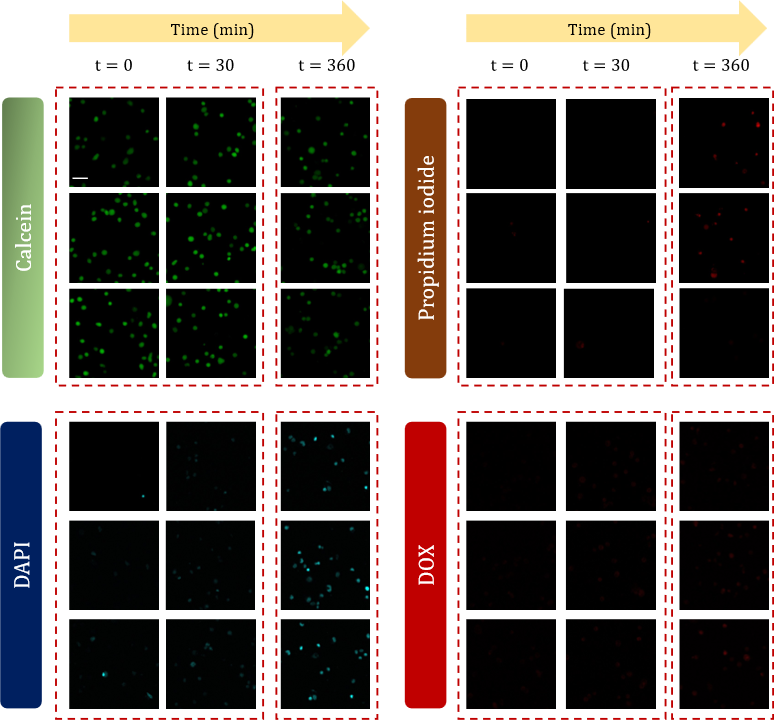}
    \caption{\textbf{Control experiments with HeLa cells.} Fluorescence emission recorded from HeLa cells exposed to different biomolecular agents and observed under corresponding fluorescence filters. Cells are trypsinized from cell culture flask, washed and resuspended in fresh cell culture medium supplemented with the biomolecular agent under investigation. These cells are then observed periodically over a period of 6 h. Permeation of biomolecular cargos is primarily through passive diffusion and is observed to relatively very slow compared to the ultrasonic-mediated transmembrane transport. Fluorescence signals under TRITC imaging filter, corresponding to propidium iodide and DOX, remains largely below detectable thresholds in the initial period till 30 mins. Scale bar indicates 50 $\mu$m. }
    \label{fig_S2}
\end{figure*}

\begin{figure*}[t]
    \centering
    \includegraphics[width=1\linewidth]{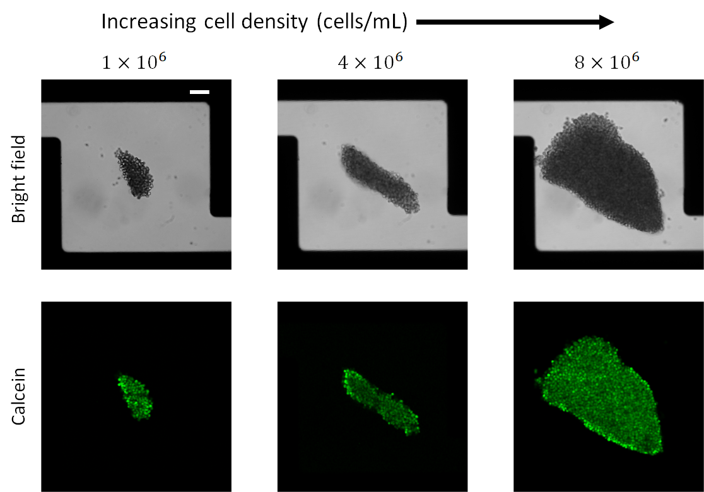}
    \caption{\textbf{Acoustic trapping in microfluidic cavity.} Experimental images of ultrasound-based trapping of HeLa cells inside the microcavity volume. Cells are stained with Calcein-AM for enhancement of visibility in fluorescence imaging filter. With increase in initial cell seeding density, the average size of the cellular aggregate grows, while the shear forces due to the shear flow are opposed by the acoustic radiation forces. Scale bar indicates 100 $\mu$m.}
    \label{fig_S3}
\end{figure*}

\begin{figure*}[t]
    \centering
    \includegraphics[width=1\linewidth]{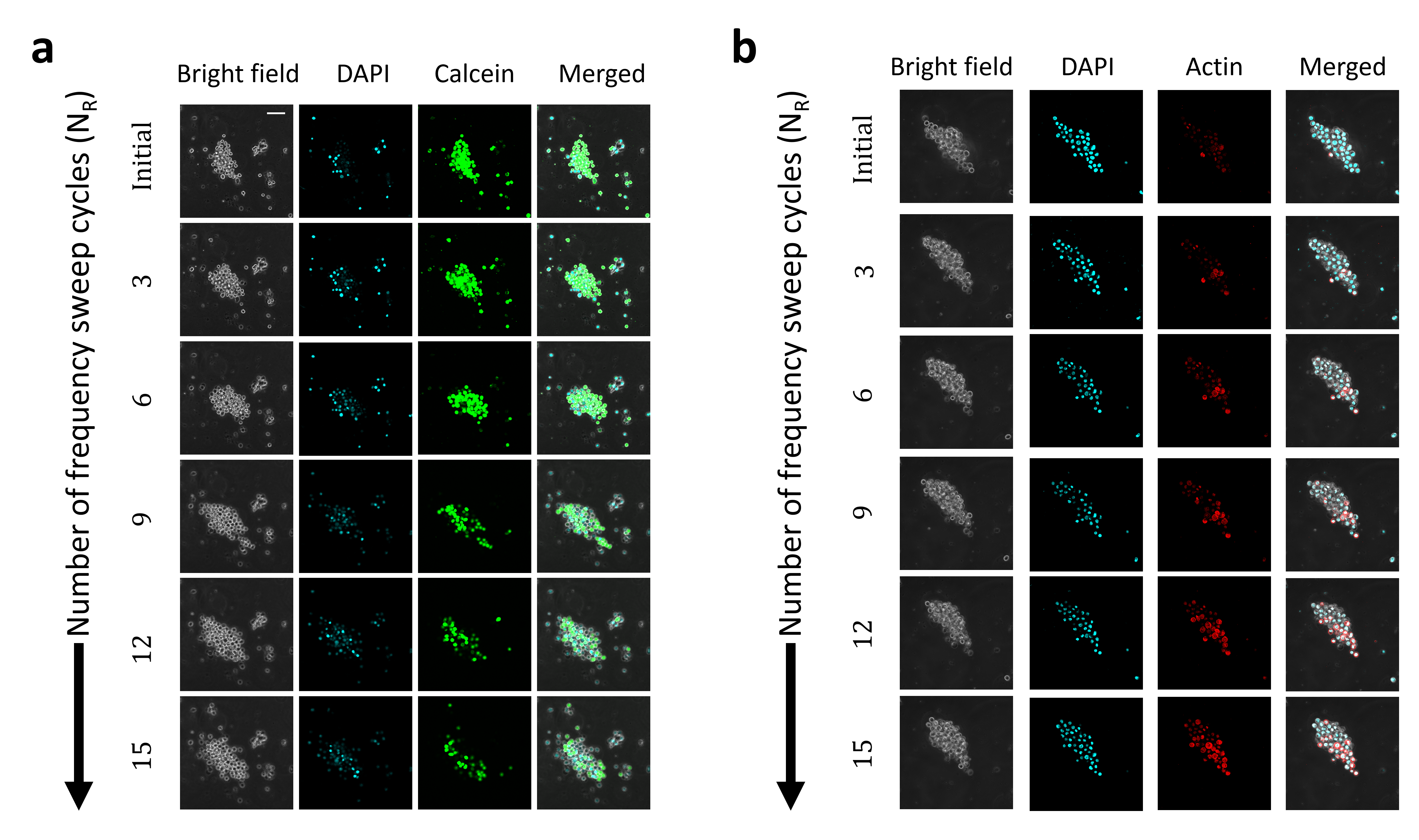}
    \caption{\textbf{Fluorescence images of intracellular delivery.} Confocal scanned images of cellular aggregate within microcavity volume and subjected to frequency sweep cycles, (a) Uptake and efflux of DAPI and fluorescent Calcein, respectively. With increase in number of frequency sweep cycles, blue fluorescence intensity corresponding to DAPI filter gradually increases, while green fluorescence intensity corresponding to FITC imaging filter exhibits a reverse trend in decreasing intensity levels, (b) Uptake of actin stain by HeLa cells; these cells were initially stained with Hoeshcht-33342 (as observed in DAPI filter) for improvement of visualization. Increase in fluorescence intensity levels in TRITC imaging filter validated the internalization of actin stain molecules. These experimental results establish the non-specific nature of transmembrane transport achieved using PAST setup. Scale bar represents 100 $\mu$m.}
    \label{fig_S4}
\end{figure*}

\begin{figure*}[t]
    \centering
    \includegraphics[width=1\linewidth]{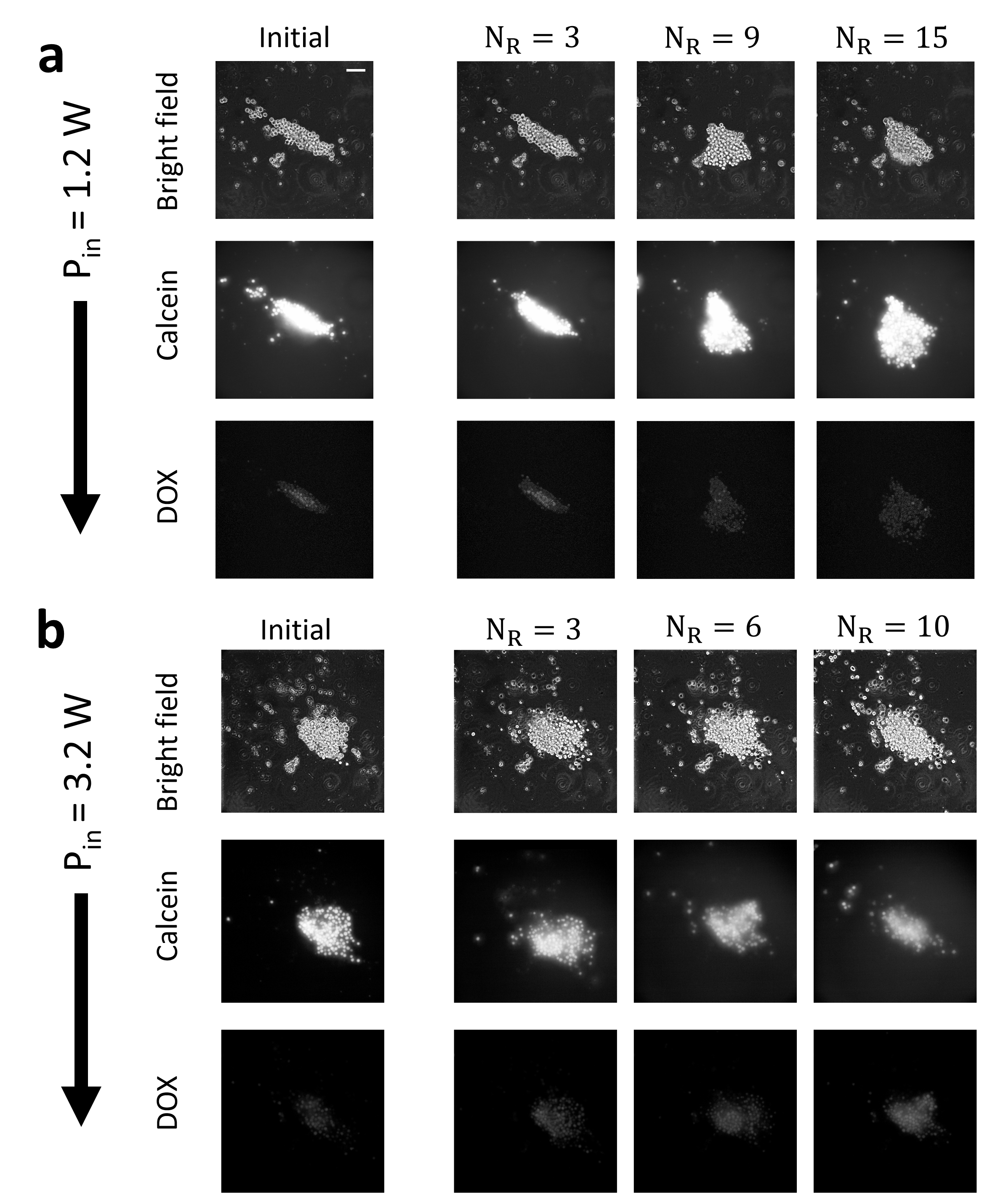}
    \caption{\textbf{Fluorescence evolution images for HeLa cells subjected to different acoustic powers.} Biomolecular transport rates of calcein and doxorubicin across the membrane of HeLa cells, subjected to two different acoustic driving powers. a. At lower driving power, $P_{in}$ = 1.2 W, the changes in fluorescence intensity levels are minimal and no efflux and uptake (for calcein and doxorubicin, respectively) can be observed even after 15 frequency sweep cycles, b. Higher driving power, $P_{in}$ = 3.2 W, in contrast, leads to significant changes in fluorescence intensity levels following 10 frequency sweep cycles. These experimental results validate higher transmembrane transport rates accompanying higher synergistic stresses imparted onto cell membranes. Scale bar represents 100 $\mu$m.}
    \label{fig_S5}
\end{figure*}

\begin{figure*}[t]
    \centering
    \includegraphics[width=1\linewidth]{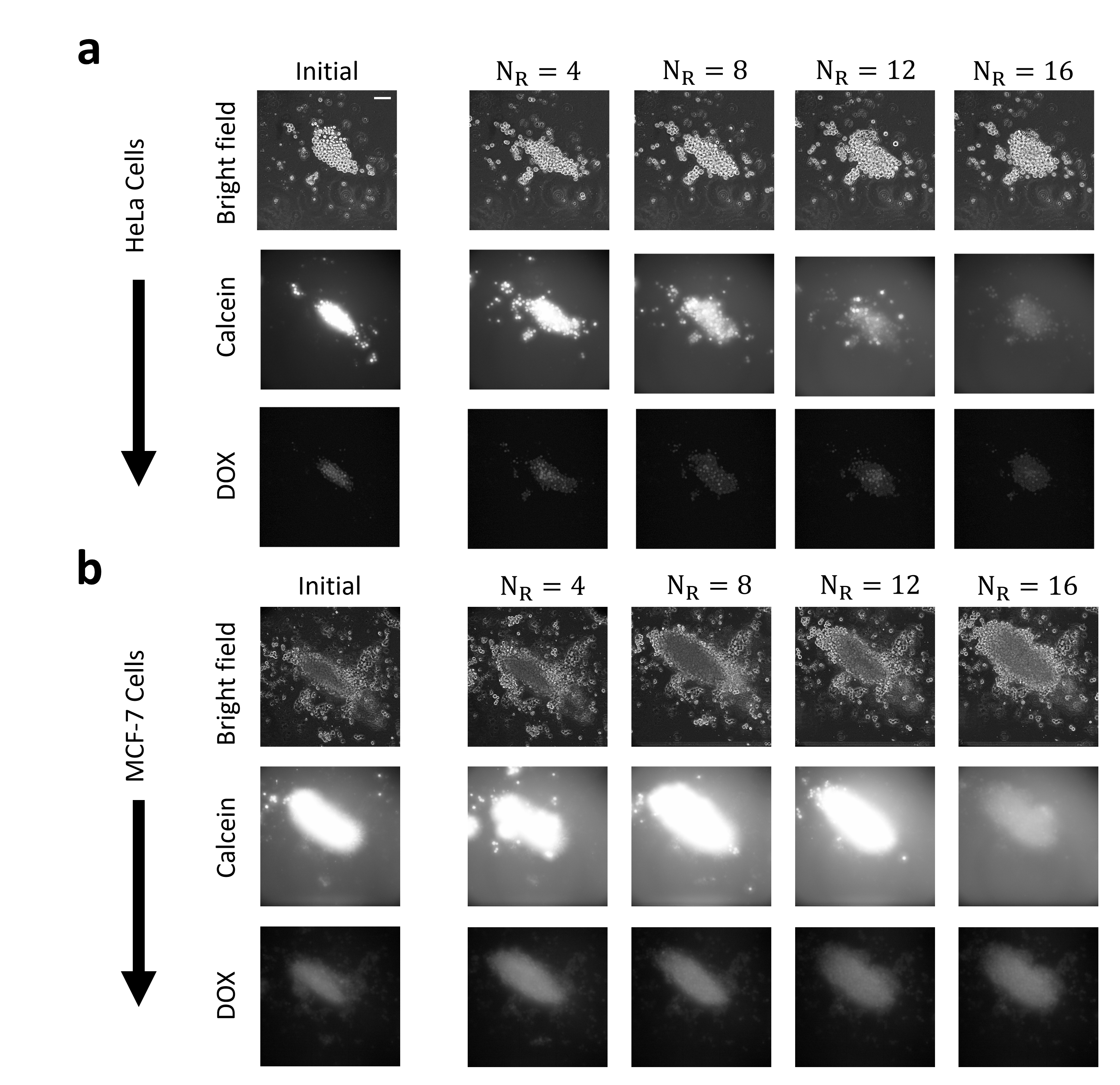}
    \caption{\textbf{Fluorescence evolution images for HeLa and MCF-7 cells.} Fluorescence intensity evolution corresponding to transport of calcein and doxorubicin for a. HeLa and b. MCF-7 cell lines. Experimental observations indicate higher uptake rates and slower efflux rates for MCF-7 cells as compared to HeLa cells; these trends could be attributed to differences in membrane potentials for these cell lines. Scale bar represents 100 $\mu$m.}
    \label{fig_S6}
\end{figure*}

\begin{figure*}[t]
    \centering
    \includegraphics[width=1\linewidth]{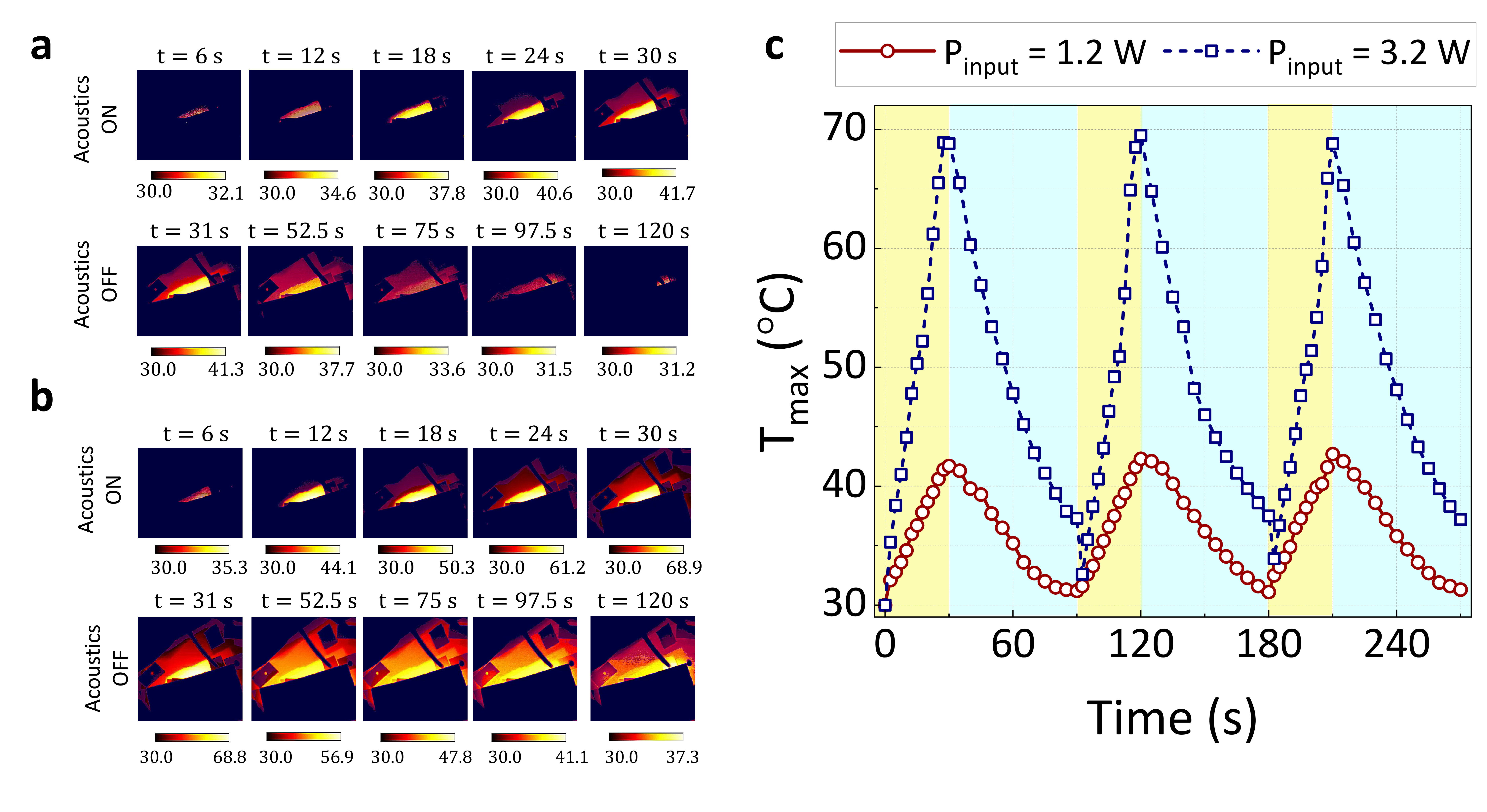}
    \caption{\textbf{Temperature changes in acoustofluidic device subjected to different acoustic driving powers.} Experimentally obtained infrared images for temperature changes in the acoustofluidic device subjected to, a. $P_{in}$ = 1.2 W, b. $P_{in}$ = 3.2 W. For each experimental run, the device is actuated for a period of $\tau_{on}$ = 30 seconds, accompanied by frequency sweep cycle, and followed by a resting period of $\tau_{off}$ = 90 seconds, c. Characterization of maximum temperature rise in the acoustofluidic device subjected to different acoustic powers. Maximum temperature corresponding to higher driving powers is significantly high, thus indicating plausible detrimental effects on cellular viability. In addition, the device does not return to its baseline temperature during the resting period, leading to a progressive rise in average temperature over multiple frequency sweep cycles.}
    \label{fig_S7}
\end{figure*}

\begin{figure*}[t]
    \centering
    \includegraphics[width=1\linewidth]{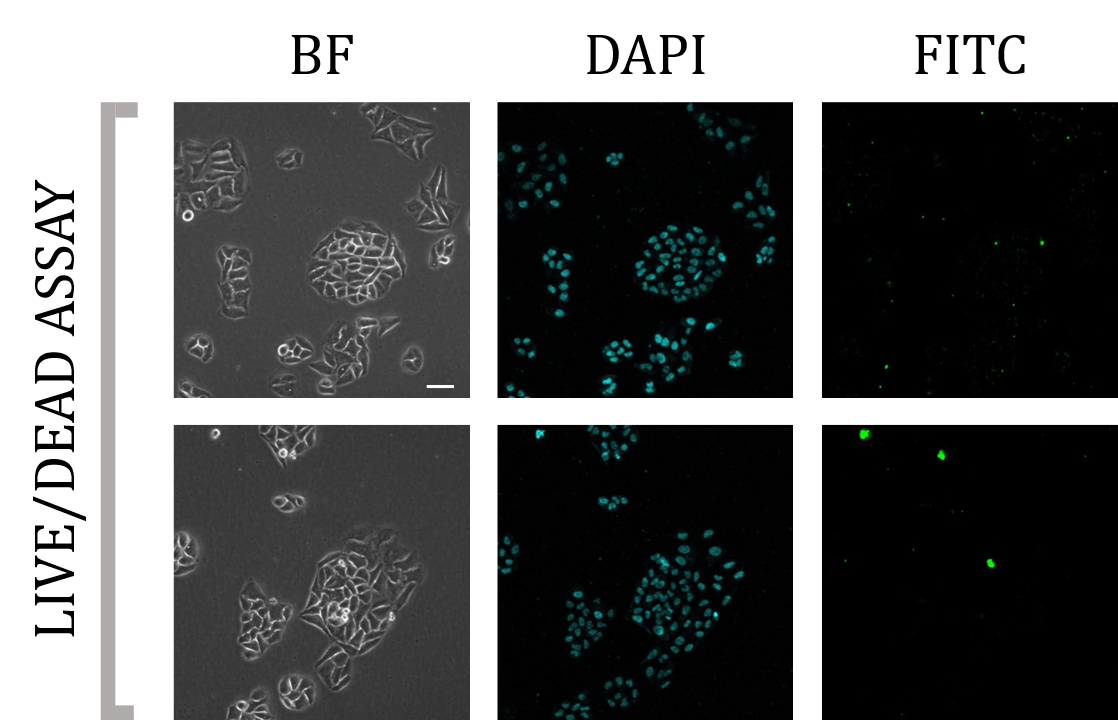}
   \caption{\textbf{Live/dead assay of ultrasonically treated cells.} Live/dead assay performed on ultrasonically treated cells after a period of 72 h and visualized under a confocal microscope. The cells are extracted and seeded onto a poly-L-lysine coated petri dish following acoustic exposure. After 72 h, the cells exhibit healthy proliferation and adherence. Upon staining, blue fluorescence emitted in DAPI imaging filter corresponds to nuclear staining of all cells within field-of-view, while green fluorescence emitted in FITC imaging filter corresponds to cells with permanently compromised cell membranes. The percentage of dead cells is observed to be minimal with respect to the total population of cells that have been ultrasonically treated and seeded, thus confirming biocompatibility of the proposed methodology. Scale bar indicates 100 $\mu$m.}

    \label{fig_S8}
\end{figure*}

\begin{figure*}[t]
    \centering
    \includegraphics[width=1\linewidth]{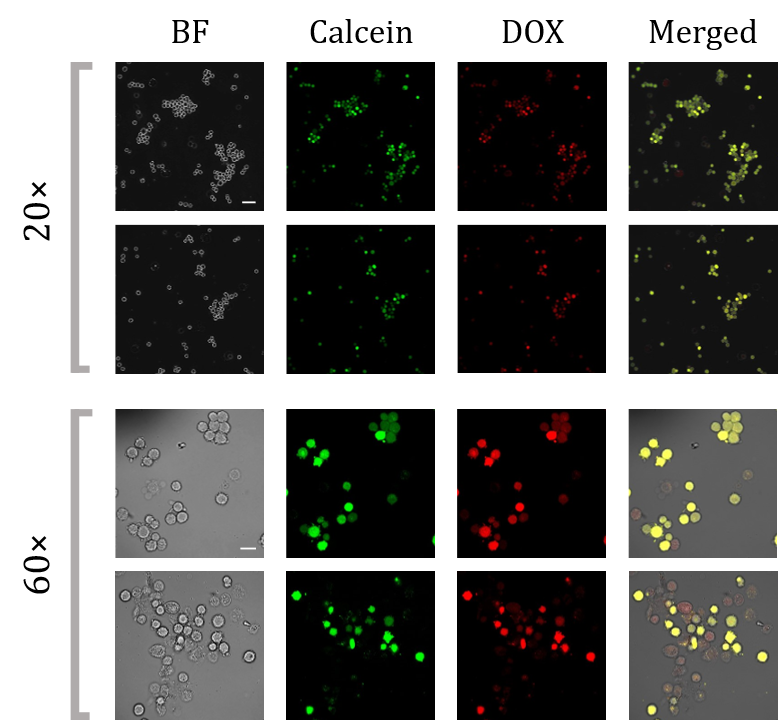}
   \caption{\textbf{Confocal scanned images of acoustically treated cells post ultrasonic exposure and seeding.} Experimentally observed fluorescence images of ultrasonically treated cells after a period of 24 h, post acoustic exposure to dynamic frequency alterations. Cells are extracted from the acoustofluidic device onto a poly-L-lysine coated petri dish, pre-filled with a small volume of cell culture medium to assist cell adherence and proliferation. Images are visualized under (a) 20$\times$ and (b) 60$\times$ magnification scales. Retention of doxorubicin within the intracellular volume is validated through the colocalized fluorescence emission within the region of cells. The cells also retain their spherical shapes and do not adhere to the poly-L-lysine coated petri dish, indicating loss of cellular viability due to intracellular permeation of doxorubicin drug molecules. Scale bar represents 100 $\mu$m in (a) and 20 $\mu$m in (b). }

    \label{fig_S9}
\end{figure*}

\begin{figure*}[t]
    \centering
    \includegraphics[width=1\linewidth]{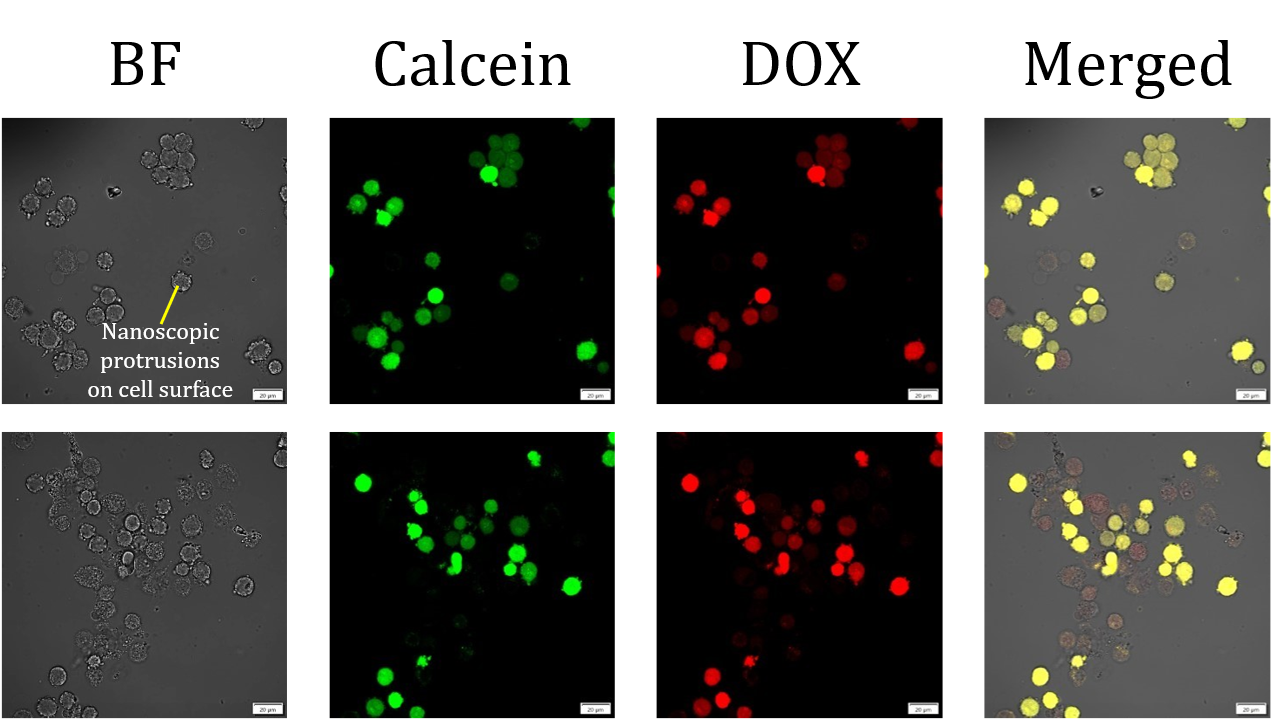}
   \caption{\textbf{Ultrasonically treated cells visualized on a confocal microscope.} Magnified images of HeLa cells treated under dynamically changing excitation frequency conditions in acoustofluidic device under a confocal microscope. The images reveal formation of nanoscopic protrusion-like structures on the surface of ultrasonically treated cells. This is likely due to bleb formation arising out of synergistic membrane and underlying cytoskeleton disruption, followed by repairing mechanisms and recovery of cellular homeostasis.  Scale bar represents 20 $\mu$m. }
    \label{fig_S10}
\end{figure*}

\begin{figure*}[t]
    \centering
    \includegraphics[width=1\linewidth]{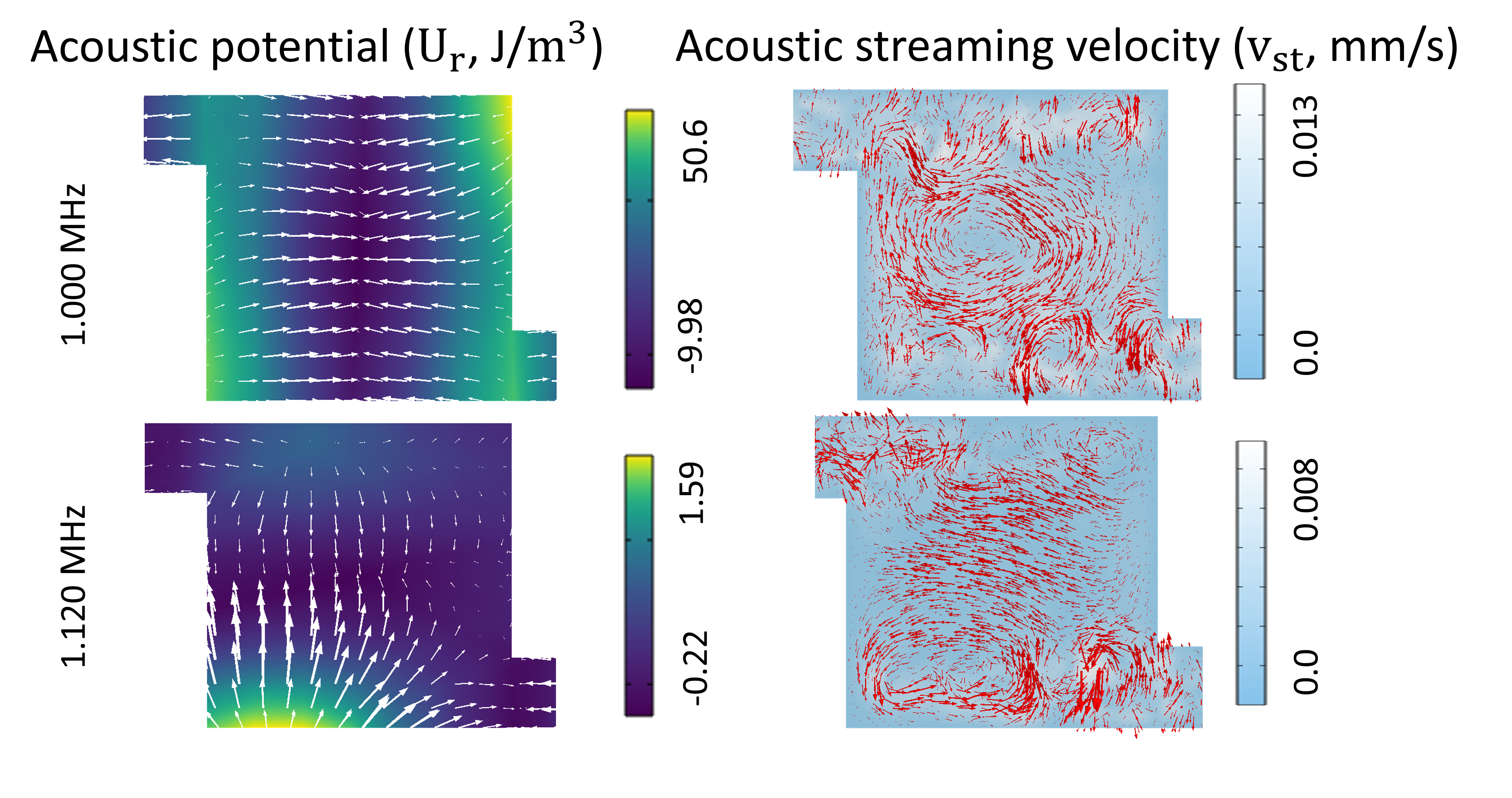}
    \caption{\textbf{Acoustic potential and streaming distributions.} Simulated acoustic potential and streaming fields at two different actuation frequencies of 1.000 MHz and 1.120 MHz, with superimposed arrows indicating primary acoustic radiation force (ARF)(white) and acoustic streaming velocity (ASV)(red), respectively. Frequency modulation induces periodic changes in acoustic landscape.}
    \label{fig_S11}
\end{figure*}

\begin{figure*}[t]
    \centering
    \includegraphics[width=1\linewidth]{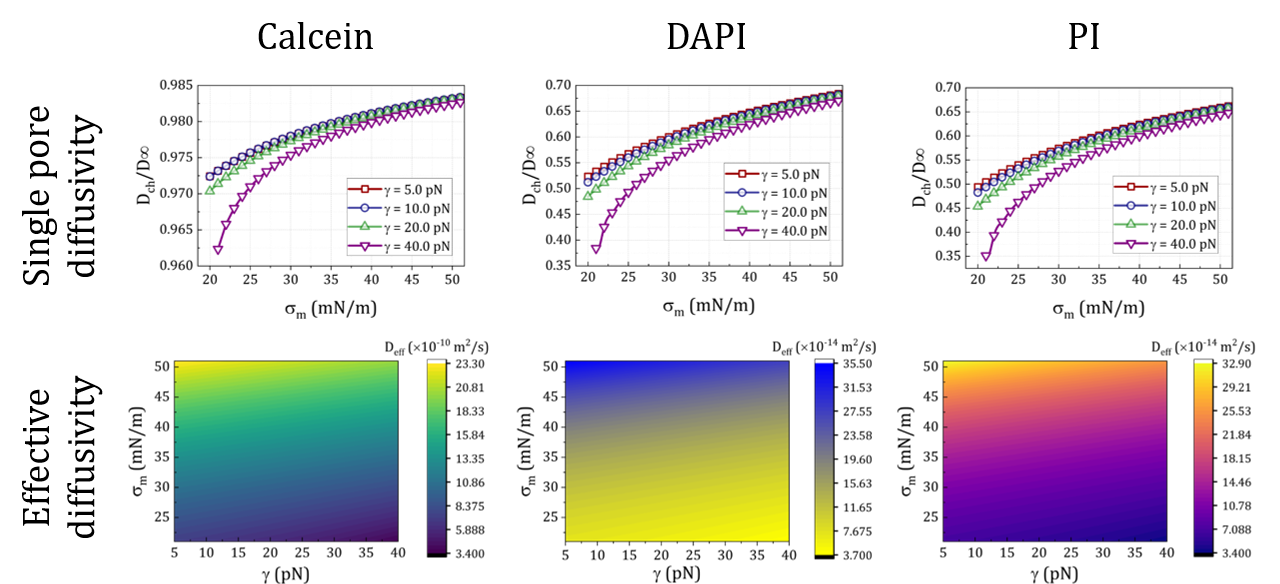}
   \caption{\textbf{Theoretical scaling of enhanced diffusivity through permeabilized cell membrane.} Theoretical scaling of ultrasound-assisted permeabilization and enhanced transport across the cell membrane for different biomolecular species. The plots on the top row illustrate the hindered diffusivity of different species through a pore formed on the cell membrane as compared to diffusivity in unbounded fluid medium. An increase in edge tension of the membrane makes it energetically difficult to retain stable pores on the membrane surface, thus reducing diffusivity. The contour plots on the bottom row illustrate the effective diffusivity through cell membranes as a function of edge tension and applied membrane tension. Increasing applied membrane tension favors formation of pores, thus enhancing intracellular transport. }

    \label{fig_S12}
\end{figure*}

\end{document}